\documentstyle[12pt]{article}
\catcode`@=11
\newcount\@tempcntc
\def\@citex[#1]#2{\if@filesw\immediate\write\@auxout{\string\citation{#2}}\fi
  \@tempcnta\z@\@tempcntb\m@ne\def\@citea{}\@cite{\@for\@citeb:=#2\do
    {\@ifundefined
       {b@\@citeb}{\@citeo\@tempcntb\m@ne\@citea\def\@citea{,}{\bf
?}\@warning
       {Citation `\@citeb' on page \thepage \space undefined}}%
    {\setbox\z@\hbox{\global\@tempcntc0\csname
b@\@citeb\endcsname\relax}%
     \ifnum\@tempcntc=\z@ \@citeo\@tempcntb\m@ne
       \@citea\def\@citea{,}\hbox{\csname b@\@citeb\endcsname}%
     \else
      \advance\@tempcntb\@ne
      \ifnum\@tempcntb=\@tempcntc
      \else\advance\@tempcntb\m@ne\@citeo
      \@tempcnta\@tempcntc\@tempcntb\@tempcntc\fi\fi}}\@citeo}{#1}}
\def\@citeo{\ifnum\@tempcnta>\@tempcntb\else\@citea\def\@citea{,}%
  \ifnum\@tempcnta=\@tempcntb\the\@tempcnta\else
   {\advance\@tempcnta\@ne\ifnum\@tempcnta=\@tempcntb \else
\def\@citea{--}\fi
    \advance\@tempcnta\m@ne\the\@tempcnta\@citea\the\@tempcntb}\fi\fi}
\catcode`@=12
\def\theequation{\arabic{section}.\arabic{equation}}
\begin{document}

\begin{flushright}
ZTF/95--06\\
October 1995
\end{flushright}

\begin{center}
{\Large{\bf Probing lepton-number/flavour-violation in}}\\[0.3cm]
{\Large{\bf semileptonic {$\tau$} decays into two mesons}}\\[1.7cm] 
{\large A.~Ilakovac}\\[0.4cm]
{\em University of Zagreb, Faculty of Science, Department of Physics,\\ 
Bijeni\v cka 32, 10000 Zagreb, Croatia}\\
(Received 27 October 1995)
\end{center}
\vskip1.7cm
\centerline{\bf ABSTRACT}
\vskip0.75cm
The evaluation, systematic
analysis and numerical study of the semileptonic $\tau$-lepton decays 
with two mesons in the final
state has been made in the frame of the 
standard model extended by right handed
neutrinos. In the analysis, heavy-neutrino nondecoupling effects,
finite quark masses, quark and meson mixings, finite widths of vector
mesons, chiral symmetry breakings in vector-meson--pseudoscalar-meson
vertices and effective Higgs-boson--pseudoscalar-meson couplings 
have been included.
Numerical estimates reveal that the decays $\tau^-\to e^-\pi^-\pi^+$,
$\tau^-\to e^-K^-K^+$ and $\tau^-\to e^-K^0\bar{K}^0$ have branching ratios 
of the order of $10^{-6}$, close to present-day experimental sensitivities.
\\

\noindent
PACS number(s): 13.35.Dx, 14.60.St, 12.39.Fe

\newpage
\section{Introduction}
\setcounter{equation}{0}
\indent

The neutrinoless $\tau$-lepton decays belong to the family of phenomena
which, if experimentaly confirmed, would unambiguously show that there
exists physics beyond the standard model (SM). Specifically, the lepton
sector would have to be modified. In the SM, these decays are forbidden, due to
the fact the SM-neutrinos $\nu_e$, $\nu_\mu$ and $\nu_\tau$ are exactly
massless, the fact which follows from the doublet nature of neutrino and
Higgs boson fields, left-handedness of the neutrinos, and chirality 
conservation. Neutrinoless $\tau$-lepton decays, if studied with
sufficient accuracy, from the experimental point of view, are very promising 
due to the large momentum transfer involved \cite{DL,HM}. 
In addition, the large mass
of the $\tau$-lepton allows many decay channels. Therefore, SM (deviations
from the SM) can be tested in a variety of ways. Experimental data on
these decays constantly improve \cite{tauexp,CLEO}. The CLEO experiment
\cite{CLEO}, has improved the 
previous upper bounds on 22 neutrinoless decay channels of the
$\tau$-lepton by almost an order of magnitude.

Neutrinoless $\tau$-lepton decays and many other lepton-number/flavour
violating decays have been studied in a number of models, e.g.
$SU(2)\times U(1)$ theories with more than one Higgs doublet \cite{SY}, 
leptoquark models \cite{PS}, $R$-parity violating supersymmetry scenarios
\cite{BGH}, superstring models with $E_6$ symmetry \cite{WUA}, left-right 
symmetric models \cite{LR} and theories containing heavy Dirac and/or Majorana 
neutrinos \cite{ZPC,SV}. Here, the models with heavy Dirac 
and/or Majorana
neutrinos will be used to estimate the processes of interest.

This paper is devoted to the analysis of semileptonic decays with two
pseudoscalar mesons in the final state, denoted by $\tau^-\to l^\mp P_1
P_2$. Together with papers \cite{NPB} and \cite{PRD}, it completes the
analysis of the lepton number/flavour violating decays
of the $\tau$-lepton reported by the CLEO collaboration \cite{CLEO}.
In addition to the heavy-neutrino nondecoupling effects 
\cite{NPB,PRD,AP_H,AP_pertu,AP}, finite quark mass
contibutions, Cabbibo-Kobayashi-Maskawa (CKM) quark mixings and meson mixings
already studied in the previous work \cite{PRD}, this analysis includes 
vector-meson--pseudoscalar meson couplings,
chiral symmetry-breaking effects, finite widths of the vector mesons and
effective Higgs-pseudoscalar couplings. The hadronic 
matrix elements are derived in a few independent ways, in order to check the 
formalism used.

For the evaluation of the leptonic part of the $\tau^-\to l^\mp P_1P_2$
matrix elements, the formalism and conventions of the 
model described in Ref.~
\cite{ZPC} are adopted. The model is based on the SM group. Its neutrino
sector is extended by the presence of a number ($n_R$) of neutral isosinglets
leading to $n_R$ heavy Majorana neutrinos ($N_j$). The quark sector of the
model retains the SM structure. In couplings of charged and neutral
current interactions, CKM-type matrices $B$ and $C$ appear 
\cite{ZPC,NPB,APetal}. These matrices satisfy a number of identities, assuring
the renormalisability of the model \cite{ZPC,APBK} and reducing 
the number of free parameters in the theory.
These identities may be used to estabilish the
relation between $B$ and $C$ matrices and neutrino masses, too. For example,
in the model with two right-handed neutrinos, $B$ and $C$ matrices read
\cite{NPB}
\begin{eqnarray}
B_{lN_1} &=& \frac{\rho^{1/4} s^{\nu_l}_L}{\sqrt{1+\rho^{1/2}}}\ ,
\qquad
B_{lN_2}\  = \frac{i s^{\nu_l}_L}{\sqrt{1+\rho^{1/2}}}\ ,
\label{BlN}
\nonumber\\
C_{N_1N_1} &=& \frac{\rho^{1/2}}{1+\rho^{1/2}}\ \sum\limits_{l=1}^{n_G}
(s^{\nu_l}_L)^2, 
\qquad 
C_{N_2N_2}\ =\ \frac{1}{1+\rho^{1/2}}\ \sum\limits_{l=1}^{n_G}
(s^{\nu_l}_L)^2,
\nonumber\\
C_{N_1N_2} &=& -C_{N_2N_1}\ =\ \frac{i\rho^{1/4}}{1+\rho^{1/2}}\
\sum\limits_{l=1}^{n_G}(s^{\nu_l}_L)^2,
\end{eqnarray}
where $\rho=m^2_{N_2}/m^2_{N_1}$, and $s^{\nu_l}_L$ are 
heavy-light neutrino mixings \cite{Lan} defined by
\begin{equation}
\label{sL}
(s^{\nu_l}_L)^2\ \equiv\ 1- \sum\limits_{i=1}^3 |B_{l\nu_i}|^2\
=\ \sum\limits_{j=1}^{n_R}|B_{lN_j}|^2.
\end{equation}
The second equation (\ref{sL}) follows from the afore-mentioned relations
for $B$ and $C$ matrices. In the theory with more than one isosinglet,
the heavy-light neutrino mixing and light-neutrino masses ($m_{\nu_l}$)
are not necessarily correlated through the traditional see-saw relation
$(s_L^{\nu_l})^2\propto m_{\nu_l}/m_M$. The $(s_L^{\nu_l})^2$ scales as
$(m_D^\dagger(m_M^{-1})^2m_D)_{ll}$ \cite{APetal,Lan}, while light-neutrino
masses depend on the matrix $m_Dm_M^{-1}m_D^T$. If the condition 
$m_Dm_M^{-1}m_D^T=0$ is fullfiled, tree-level light-neutrino masses 
are equal zero, while $(s_L^{\nu_l})^2$ can assume large values. 
The light neutrinos receive nonzero values radiatively, but for
reasonable $m_M$ values, their values are in agreement with the experimental
upper bounds \cite{ZPC}. Independence of the light-neutrino masses and
the heavy-light neutrino mixings implies that $(s_L^{\nu_l})^2$ may be
treated as free phenomenological parameters, which may be constrained by
low energy data \cite{Lan,BGKLM}. In this way, the following upper limits 
for the heavy-light neutrino mixings have been found \cite{BGKLM}:
\begin{equation}
\label{upli}
(s^{\nu_e}_L)^2, (s^{\nu_\mu}_L)^2 < 0.015,\qquad
(s^{\nu_\tau}_L)^2<0.050,
\qquad (s^{\nu_e}_L)^2(s^{\nu_\mu}_L)^2<10^{-8}.
\end{equation}
More recently, a global analysis of all available
electroweak data accumulated at the CERN Large Electron
Positron Collider (LEP) has yielded the more stringent
limits~\cite{Roulet},
\begin{equation}
\label{upli2}
(s^{\nu_e}_L)^2<0.0071,\qquad (s^{\nu_\mu}_L)^2 < 0.0014,\qquad
(s^{\nu_\tau}_L)^2<0.033 \quad (0.024\ \mbox{including LEP data}),
\end{equation}
at the $90\%$ confidence level (CL). In this paper, the limits obtained 
in the Ref.~\cite{BGKLM} will be used because the results of the
analysis in Ref.~\cite{Roulet} depend to certain extent on the CL
considered in the global analysis and on some model-dependent
assumptions \cite{NPB}. The discussion on possible theoretical 
dependence of the 
upper limits, such as those in Eqs.~(\ref{upli}) and (\ref{upli2}), 
may be found in Ref.~\cite{PRD}.

The hadronic part of the amplitudes contains matrix elements of 
quark currents between vacuum and a hadronic state. Vector and
axial-vector quark currents are identified with vector and 
pseudoscalar mesons through PCAC \cite{PCAC} and vector meson dominance 
\cite{VMD,EVMD,KRS_VMD}
relations. The scalar quark current is expressed in terms of 
pseudoscalar mesons, identifying QCD and chiral-model Lagrangian.
Intermediate vector mesons are
described by the Breit-Wigner propagators with momentum-independent width
\cite{FWWACS,Decker,Palmer}.
The vector-meson--pseudoscalar vertices are described by 
non-gauged $U(3)_L\times U(3)_R/U(3)_V$ chiral Lagrangian containing   
hidden $U(3)_{local}$ symmetry \cite{BandoPRep}, 
through which the vector mesons are
introduced. Both $U(3)_L\times U(3)_R/U(3)_V$-symmetric and more realistic
$U(3)_L\times U(3)_R/U(3)_V$-broken 
Lagrangians \cite{BandoNPB} are used in the evaluation
of the matrix elements. The gauge couplings of mesons are introduced
indirectly through the quark gauge couplings in the above mentioned 
matrix elements of quark currents.

This paper is organized as follows. In Section~2, the analytical
expressions for branching ratios of decay processes 
$\tau^-\to e^+ P_1^-P_2^-$ and $\tau^-\to e^- P_1^-P_2^+/e^-P_1^0P_2^0$ 
are derived.
Technical details are relegated to the Appendices.
Numerical results are presented in Section~3.
Conclusions are given in Section 4.

\section{{$\tau^-\rightarrow l'^\mp P_1P_2$}}
\setcounter{equation}{0}

In the model containing heavy Majorana neutrinos, 
there are two possible types of the 
semileptonic $\tau$-lepton decays into two pseudoscalar mesons
\begin{itemize}
\begin{enumerate}
\item $\tau^-\rightarrow l'^+ P_1^-P_2^-$ and
\item $\tau^-\rightarrow l'^- P_1^{Q_1}P_2^{Q_2}$, $Q_1+Q_2=0$,
\end{enumerate}
\end{itemize}
where $P_1$ and $P_2$ are pseudoscalar mesons, and $Q_1$ and $Q_2$ are their
charges. Type (1) violates both
lepton flavour and lepton number, and requires the exchange of Majorana
neutrinos; henceforth these reactions
will be refered to the {\em Majorana type}.
Type (2) violates lepton
flavour and proceeds via the exchange of Dirac or Majorana neutrinos;
the appelation {\em Dirac type} will be attributed to these decays.
Feynman diagrams 
pertinent to the Majorana-type and Dirac-type decays are given on 
Fig.~1(a) and Fig.~1(b), respectively.
As mentioned in Introduction, only the decays with 
two-pseudoscalar final states, which are currently under experimental
investigation, are considered.
The decays with other two-meson final states
could be calculated within the model, too, but 
they are phase-space 
suppressed, they haven't been experimentally searched for, and
they decay into the final states with more than 
two pseudocalar mesons. The
complete calculation of such decays is much more involved 
than for the decays with two pseudoscalar mesons in the final state
\cite{Decker}.

To start with, we consider the Majorana-type decays. At the
lowest, fourth order in the weak interaction coupling constant, only
tree diagrams contribute to the Majorana type decays. The chirality
projection operators project out the mass terms of the numerators of the
neutrino propagators. For that reason, only massive neutrinos
contribute to the $\tau^-\to l'^+P_1^-P_2^-$ amplitude. Since the
$W$-boson and heavy neutrino masses \cite{ZPC} are much larger than
the energy scale at which quarks hadronize to mesons, 
their propagators may be shrunk to points 
so as to form an effective amplitude
depending only on one space-time coordinate:
\begin{eqnarray}
\label{SMaj}
S(\tau^-\to l'^+P_1P_2)&=&\frac{-i\alpha_W^2\pi^2}{2M_W^4}
 \sum_{a,b=1}^2 V_{ud_a}^\ast V_{ud_b}^\ast
\sum_{i=1}^{n_R}\frac{B_{l'N_i}^\ast B_{\tau N_i}^\ast}{m_{N_i}}
\bar{u}_{l'}(1-\gamma_5)u_\tau\int d^4x e^{-i(p-p')x}
\nonumber\\&&
\langle P_1^-P_2^-|\bar{d}_a(x)\gamma^\mu(1-\gamma_5)u(x)
\bar{d}_b(x)\gamma_\mu(1-\gamma_5)u(x)|0\rangle,
\end{eqnarray}
where $\alpha_W=\alpha_{\rm em}/\sin^2\theta_W\approx0.0323$ is the weak
fine-structure constant, $M_W$ is $W$-boson mass,
$V_{ud_a}$ are CKM
matrix elements, $m_{N_i}$ are heavy neutrino masses and $u(x)$ and $d_a(x)$ 
are quark fields for $u$, $d$ and $s$ quark ($d_1=d$ and $d_2=s$).
A more reliable calculation would also include the QCD corrections
of four quark operators in eq. (\ref{SMaj}) (they introduce new quark
operators, and mixing of all quark operators),
along with a renormalization-group analysis
of their coefficients \cite{rengr,vqa}.
Since such refinements will not alter our conclusions concerning the
magnitude of the amplitude, they will be ignored.

The hadronic matrix element may be evaluated using vacuum saturation
approximation and PCAC. Vacuum saturation approximation \cite{vqa,DGH}
allows to split
the matrix elements involving four-quark operators into matrix elements
of two-quark operators. The two-quark operators forming axial-vector
currents may be combined into the currents 
having the same quark content as the produced pseudoscalar mesons, $P$,
$A^P_\mu(x)$.
The matrix elements of the currents $A^P_\mu(x)$ are evaluated using the
PCAC relation \cite{PCAC}:
\begin{equation}
\label{PCAC}
\langle 0|A^P_\mu(x)|P'(p_{P'})\rangle = 
\delta_{PP'}\sqrt{2}f_{P'}p^{P'}_\mu e^{-ip_{P'}x},    
\end{equation}
where $f_{P'}$ is the decay constant of pseudoscalar meson $P'$.
The Kronecker symbol $\delta_{PP'}$ assures that the matrix elements
(\ref{PCAC}) give the nonzero result only if the final state quantum
numbers match those of the axial-vector current. Following the above
procedure, one obtains the expression for the generic matrix element of 
$\tau^-\to l'^+P_1^-P_2^-$ process
\begin{eqnarray}
T(\tau^-\to l'^+P_1^-P_2^-)&=&-\frac{i8\alpha_W^2\pi^2}{3}
V_{ud_a}^\ast V_{ud_b}^\ast\frac{f_{P_1}f_{P_2}}{M_W^4}
\sum_{i=1}^{n_R}B_{l'N_i}^\ast B_{\tau N_i}^\ast\frac{1}{m_{N_i}}
\nonumber\\&&
\times(p_{P_1}p_{P_2})\bar{u}_{l'}(1-\gamma_5)u_\tau.
\end{eqnarray}
The corresponding branching ratio reads
\begin{equation}
B(\tau^-\to l'^+P_1^-P_2^-)\ =\ 
S\;\frac{\alpha_W^4\pi(f_{P_1} f_{P_2})^2}{36\Gamma_\tau m^3M_W^{10}}
|V_{ud_a}V_{ud_b}|^2 
\bigg|\sum_{i=1}^{n_R}B_{l'N_i}B_{\tau N_i}\frac{M_W}{m_{N_i}}\bigg|^2
\int_{(m_1+m_2)^2}^{(m-m')^2}dt\:\omega,\qquad
\end{equation}
where $S$ is the statistical factor, equal to $1/2$ if two equal
pseudoscalars appear in the final state, and 
$\omega$ is a phase-space integral of the Mandelstam-variables dependent
part of the square of the amplitude which is defined  
in Appendix C.

Now we turn to the Dirac-type decays. The scattering matrix element
of $\tau^-\rightarrow l'^- P_1P_2$ receives contributions
from $\gamma$--exchange graphs, $Z$-boson--exchange graphs, box graphs,
Higgs-boson($H$)--exchange graphs and $W^+$-boson-$W^-$-boson--exchange graphs,
\begin{eqnarray}
\label{Stot}    
S(\tau^-\to l'^- P_1P_2)&=&S_\gamma(\tau^-\to l'^- P_1P_2)
+S_Z(\tau^-\to l'^-P_1P_2)+S_{Box}(\tau^-\to l'^- P_1P_2)\nonumber\\
&&+S_{H}(\tau^-\to l'^-P_1P_2)+S_{W^-W^+}(\tau^-\to l'^-P_1P_2).\qquad 
\end{eqnarray}
The $\gamma$, $Z$-boson  and Higgs-boson amplitudes factorize 
into leptonic vertex
corrections and hadronic pieces.
The loop integrations are straightforward.
The hadronic parts of the $\gamma$- and $Z$-boson amplitudes consist of
the vacuum-to-vector-meson matrix element of the local vector 
and axial-vector quark current (only vector quark currents have
nonzero contributions, since
only vector mesons decay into the two-pseudoscalar-meson state), 
a propagator of the 
vector meson and the vector-meson-$P_1$-$P_2$ vertex. The hadronic part
of the $H$ amplitude contains vacuum-to-$P_1$-$P_2$ matrix element of
the local scalar quark current.
Exploiting translation invariance, the phases that describe the
motion of the meson(s) formed in a vacuum-to-hadron matrix element
may be isolated. Therefore, only the 
space-time independent hadronic matrix elements remain.
These phases assure four-momentum conservation. The $\gamma$, 
$Z$-boson and Higgs-boson
amplitudes read
\begin{eqnarray}
\label{SgZH}
S_\gamma(\tau^-\to l'^- P_1P_2) &=&
(2\pi)^4\,\delta^{(4)}(p-p'-p_1-p_2)\:
\nonumber\\&&
\sum_{\tilde{V}^0}
T^\mu_\gamma(\tau\to l'\tilde{V}^0)\; iS_{\tilde{V}^0,\mu\nu}(q)\; 
         T^\nu(\tilde{V}^0\to P_1P_2)
\nonumber\\
S_Z(\tau^-\to l'^- P_1P_2) &=&
(2\pi)^4\,\delta^{(4)}(p-p'-p_1-p_2)\:
\nonumber\\&&
\sum_{\tilde{V}^0}
T^\mu_Z(\tau\to l'\tilde{V}^0)\; iS_{\tilde{V}^0,\mu\nu}(q)\; 
         T^\nu(\tilde{V}^0\to P_1P_2)
\nonumber\\
S_H(\tau^-\to l'^- P_1P_2) &=& 
(2\pi)^4\,\delta^{(4)}(p-p'-p_1-p_2)\:
T_H(\tau\to l'P_1P_2),
\label{SGZH}
\end{eqnarray}
where $p$,$p'$,$p_1$ and $p_2$ are the four-momenta of $\tau$, $l'$, 
$P_1$ and $P_2$, respectively, 
$\sum_{\tilde{V}^0}$ is a sum over
vector mesons that appear simultaneously in $T^\mu_{\gamma,Z}$ and
$T^\nu(\tilde{V}^0\to P_1P_2)$ amplitudes,
$S_{\tilde{V}^0,\mu\nu}(q)$ is a constant-width 
Breit-Wigner
propagator \cite{FWWACS,Decker,Palmer} of the vector meson $\tilde{V}_0$,
\begin{equation}
\label{BWP}
S_{\tilde{V}^0,\mu\nu}(q)=\frac{-g_{\mu\nu}
+\frac{q_\mu q_\nu}{M_{\tilde{V}^0}^2}}
{q^2-M_{\tilde{V}^0}^2+iM_{\tilde{V}^0}\Gamma_{\tilde{V}^0}},
\end{equation}
$T^\nu(\tilde{V}^0\to P_1P_2)$ multiplied by the $\tilde{V}$
polarisation vector, $\varepsilon^{\tilde{V}^0}_\mu(q)$,
gives a $\tilde{V}^0-P_1-P_2$ vertex, 
which may be read from the Lagrangians (\ref{LchSIM}) and (\ref{LchBR}),
$T^\mu_{\gamma,Z}(\tau\to l'\tilde{V}^0)$ are 
$\gamma$- and $Z$-
parts of the
$T$-matrix elements for the $\tau\to l'\tilde{V}^0$ reaction 
\cite{NPB}, from which a polarization vector of the 
$\tilde{V}^0$ meson is removed, 
\begin{eqnarray}
\label{Tg_V}
T_\gamma(\tau\to l'\tilde{V}^0)&=&T^\mu_{\gamma}(\tau\to l'\tilde{V}^0)
\varepsilon^{\tilde{V}^0}_\mu(q)
\ =\  -ieL_\gamma^\mu\langle \tilde{V}^0|j^{em}_\mu(0)|0\rangle
\nonumber\\
&\equiv &\frac{i\alpha^2_W s_W^2}{4M_W^2}\,
\bar{u}_{l'}\Big[F_\gamma^{\tau l'}(\gamma^\mu-
   \frac{q^\mu\not\!q}{q^2})(1-\gamma_5)
\nonumber\\
&&-G_\gamma^{\tau l'}\frac{i\sigma^{\mu\nu}q_\nu}{q^2}
   (m (1+\gamma_5)+m'(1-\gamma_5))\Big]u_\tau
\nonumber\\
&&\times \langle \tilde{V}^0|\frac{2}{3}\bar{u}(0)\gamma_\mu u(0)
           -\frac{1}{3}\bar{d}(0)\gamma_\mu d(0)
-\frac{1}{3}\bar{s}(0)\gamma_\mu s(0) |0\rangle,\qquad
\qquad
\\
\label{TZ_V}
T_Z(\tau\to l'\tilde{V}^0)&=&T^\mu_Z(\tau\to l'\tilde{V}^0)
\varepsilon^{\tilde{V}^0}_\mu(q)
\ =\ \frac{-ig_W}{4c_W}L_Z^\mu\langle\tilde{V}^0|V^Z_\mu(0)-A^Z_\mu(0)|0\rangle
\nonumber\\
&\equiv &\frac{i\alpha^2_W}{16M_W^2}\, F_Z^{\tau l'}\,
\bar{u}_{l'}\gamma^\mu(1-\gamma_5)u_\tau
\nonumber\\
&&\times  \Bigg( \langle \tilde{V}^0|\bar{u}(0) \gamma_\mu \Big(1-\gamma_5
-\frac{8}{3}s_W^2\Big) u(0)|0\rangle\nonumber\\
&&-\langle \tilde{V}^0|\bar{d}(0)\gamma_\mu\Big( 1-\gamma_5
-\frac{4}{3}s_W^2\Big)d(0)|0\rangle \nonumber\\
&&-\langle \tilde{V}^0 | \bar{s}(0)\gamma_\mu\Big( 1-\gamma_5
-\frac{4}{3}s_W^2\Big)s(0) |0\rangle\Bigg), \label{SZ}
\end{eqnarray}
and $T_H(\tau\to l'P_1P_2)$ is the T-matrix element of the $\tau\to
P_1P_2$ reaction,
\begin{eqnarray}
\label{TH_V}
T_H(\tau\to l'P_1P_2)&=& 
\frac{-i\alpha_W^2}{8M_H^2M_W^2}
\big(m\bar{u}_{l'}(1+\gamma_5)u_\tau\: F_H^{\tau l'}
+m'\bar{u}_{l'}(1-\gamma_5)u_\tau\: G_H^{\tau l'}\big)
\nonumber\\
&&\langle P_1P_2| m_u\bar{u}(0)u(0)
               +m_d\bar{d}(0)d(0)+m_s\bar{s}(0)s(0)|0\rangle\qquad
\end{eqnarray}
In Eqs.~(\ref{BWP}--\ref{TH_V}) $m_\tau$, $m'$, 
$M_H$, $m_u$, $m_d$ and $m_s$ are masses of the $\tau$, $l'$, Higgs
boson, $u$, $d$ and $s$ quarks respectively, $s_W=\sin\theta_W$ is sine
of the Weinberg angle, $L_\gamma^\mu$
and $L_Z^\mu$ represent $\tau\to l'\gamma$ and $\tau\to l'Z$ loop functions,
respectively, multiplied by corresponding gauge-boson propagators, 
$j^{em}_\mu(0)$ is quark electromagnetic current,
and $V^Z_\mu(0)$ and $A^Z_\mu(0)$ are vector and axial-vector quark 
currents for quark--$Z$-bozon interaction.
The loop form factors $F_H^{\tau l'}$ and  $G_H^{\tau l'}$ may be found
in Appendix~B  and
$F_\gamma^{\tau l'}$ and $F_Z^{\tau l'}$ 
in Eq.~(\ref{SGZH}) in Ref.~\cite{NPB}.

The box and $W^+$-$W^-$ diagrams are more involved as they contain 
bilocal hadron currents. In the case of the box diagram, the bilocality
problem can be overwhelmed since the two $W$-bosons
in the loop assure the high virtualities of the loop momenta. That 
allows one to approximate the loop-quark propagator with the free quark
propagator, and to replace the bilocal vector and axial-vector 
current operators with the local
ones \cite{PRD}. As in $\gamma$- and $Z$- amplitudes, only the vector
quark current operators contribute, giving rise to the vector mesons,
which decay into the two-pseudoscalar-meson final state.
In this way one arrives at the following 
expression for
the box $S$-matrix element
\begin{eqnarray}
\label{SBox}
S_{Box}(\tau^-\to l'^- P_1P_2) &=&
(2\pi)^4\,\delta^{(4)}(p-p'-p_1-p_2)\:
\nonumber\\ &&
\sum_{\tilde{V}^0}
T^\mu_{Box}(l\to l'\tilde{V}^0)\; iS_{\tilde{V}^0,\mu\nu}(q)\; 
T^\nu(\tilde{V}^0\to P_1P_2),
\end{eqnarray}
where $T^\mu_{Box}(l\to l'\tilde{V}^0)$ is the box part of the 
$T$-matrix element for 
the process $l\to l'\tilde{V}^0$ \cite{NPB}, from which the polarization 
vector of the vector meson, $\tilde{V}^0$, is removed,
\begin{eqnarray}
\label{TBox_V}
T_{Box}(l\to l'\tilde{V}^0)&=&T^\mu_{Box}(l\to l'\tilde{V}^0)
\varepsilon^{\tilde{V}^0}_\mu(q)
\ =\ L_{Box,uu}^\mu\langle \tilde{V}^0|V^{Box,uu}_\mu(0)
                -A^{Box,uu}_\mu(0)|0\rangle
\nonumber\\&&
 -\sum_{d_{a,b}=d,s}L_{Box,d_ad_b}^\mu\langle \tilde{V}^0|V^{Box,d_ad_b}_\mu(0)
                -A^{Box,d_ad_b}_\mu(0)|0\rangle
\nonumber\\
&=&\frac{i\alpha^2_W}{16M_W^2}\,
\bar{u}_{l'}\gamma_\mu (1-\gamma_5)u_\tau
\Big[F_{Box}^{\tau l'uu}
    \langle \tilde{V}^0|\bar{u}(0)\gamma^\mu (1-\gamma_5)u(0)|0\rangle
\nonumber\\
&&-\sum_{d_{a,b}=d,s}F_{Box}^{\tau l'd_ad_b}
    \langle \tilde{V}^0|\bar{d}_a(0)\gamma^\mu (1-\gamma_5)d_b(0)|0\rangle\Big],
\end{eqnarray}
where $L_{Box,qq'}$ are box loop functions, and $V^{Box,qq'}_\mu(0)$ and
$A^{Box,qq'}_\mu(0)$ are the corresponding vector and axial-vector quark
currents in an $\tau\to l'\bar{q}q'$ amplitude.
The loop form factors 
$F_{Box}^{\tau l'd_ad_b}$ and $F_{Box}^{\tau l'uu}$ are defined
in Ref.~\cite{PRD}. 

As in the $\tau^-\to l^-P_1^-P_2^-$ amplitude,
the W-bosons in the $W^+$-$W^-$-exchange diagram may be shrunk
to points. So, an effective amplitude depending on two space coordinates
is formed. The chiral projection operators extract the momentum
dependent parts of the numerators of the neutrino propagators, so that
both heavy and light neutrinos contribute. The heavy-neutrino propagators
could also be shrunk to a point, and, therefore, the corresponding amplitudes 
depend on one space-time coordinate. By contrast, light-neutrino
contibutions cannot be reduced from the bilocal to a local form. To
enable the comparison of contributions of heavy and light neutrinos, all
contributions to the transition matrix element are written in their
bilocal form,
\begin{eqnarray}
\label{SWpWm}
S(\tau^-\to l'^- P_1 P_2)\hspace{-10pt}&=&
\frac{i\alpha_W^2\pi^2}{2M_W^4}
\sum_{d_{a,b}=d,s}V^*_{ud_a}V_{ud_b}
\sum_{i=1}^{n_R}B_{l'N_i}B^*_{\tau N_i}
\int d^4x d^4y\frac{d^4l}{(2\pi)^4}
\nonumber\\ &&
\times e^{i(l-p)x+i(p'-l)y}
\bar{u}_{l'}\gamma_\nu(\frac{\not\!l}{l^2}+\frac{\not\!l}{m_N^2})
\gamma_\mu(1-\gamma_5)u_\tau
\nonumber\\ &&
\times\langle P_1 P_2| \bar{u}(y)\gamma_\nu(1-\gamma_5)d_b(y)
\bar{d}_a(x)\gamma_\mu(1-\gamma_5)u(x)|0\rangle.
\end{eqnarray}
As $l^2\leq m_\tau^2$ and the lightest heavy-neutrino mass exceeds
$100\: GeV$ \cite{ZPC}, the local (heavy-neutrino) terms are supressed
at least by factor $10^{-4}$ relatively to the nonlocal (light-neutrino)
terms. Therefore, one can safely neglect them.

The amplitudes (\ref{SgZH}), (\ref{SBox}) and (\ref{SWpWm}) comprise
three types of hadronic matrix elements: 
$\langle\tilde{V}^0|\bar{q}(0)\gamma_\mu q(0)|0\rangle$,
$\langle P_1P_2|m_q\bar{q}(0)q(0)|0\rangle$ and 
$\langle P_1P_2|\bar{u}(x)\gamma_\mu d_a(x)\bar{d}_b(y)\gamma_\nu
u(y)|0\rangle$. 

The evaluation of the $\langle\tilde{V}^0|\bar{q}(0)\gamma_\mu
q(0)|0\rangle$ matrix element
proceeds as follows.
The two-quark operator $\bar{q}(0)\gamma_\mu q(0)$ is 
expressed in terms of vector
currents, $V_\mu$, having the same quark content as the produced vector
mesons, $\tilde{V}^0$. Exploiting the vector-meson dominance
relation \cite{VMD},
correlating a vector-meson field $\tilde{V}_\mu(x)$ and vector current, 
$V_\mu$, having the same quark content as $\tilde{V}_\mu(x)$,
\begin{equation}
V_\mu^{\tilde{V}}(x)=\frac{m_{\tilde{V}}^2}{\sqrt{2}\gamma_{\tilde{V}}}
\tilde{V}_\mu(x),\label{EVMD}
\end{equation}
one arrives at the expression 
\begin{equation}
\langle 0|V^{\tilde{V}'}_\mu(x)|\tilde{V}^0(p_{\tilde{V}^0})\rangle=
\delta_{\tilde{V}'\tilde{V}^0}
\frac{m_{\tilde{V}^0}^2}{\sqrt{2}\gamma_{\tilde{V}^0}}
  \varepsilon_{\tilde{V}^0\mu}(p_{\tilde{V}^0}, \lambda_{\tilde{V}^0})
e^{-ip_{\tilde{V}^0}x}.
\end{equation}
The Kronecker symbol $\delta_{\tilde{V}'\tilde{V}^0}$, assures
that the matrix elements give non-zero contributions only if the
vector-meson
quantum numbers match those of the vector current.

The $\langle P_1P_2| \sum_{q=u,d,s}m_q\bar{q}(0)q(0)|0\rangle $
matrix elements may be evaluated 
comparing  
the quark sector of the SM Lagrangian, and the corresponding
effective chiral Lagrangian, contained in the first and
second curly bracket of Eq. (\ref{LchSIM}), one obtains the expression
for the scalar two-quark current in terms of pseudoscalar fields \cite{BBG}
\begin{equation}
\label{q_scalar}
\bar{q}(x)^i q(x)^j=-\frac{1}{4}f_\pi^2\, r\:
\bigg[U(x)+U(x)^\dagger\bigg]^{ij},
\end{equation}
where $U(x)=exp(2i\pi(x)/f_\pi)$, $\pi(x)=T^a\pi^a(x)$,
$\pi^a(x)$ are pseudoscalar meson fields,
$T^a=\lambda^a/2$, $\lambda^a$ are the Gell-Mann matrices and 
\begin{equation}
\label{r}
r=\frac{2m_\pi^2}{m_d+m_u}=\frac{2m_{K^0}^2}{m_d+m_s}
 =\frac{2m_{K^+}^2}{m_u+m_s}.
\end{equation}
Exploiting Eq.~(\ref{q_scalar}), one can write 
the $H-\bar{q}-q$ part of the Yukawa Lagrangian
in terms of pseudoscalar fields
\begin{eqnarray}
{\cal L}_{H\bar{q}q}&=&-\frac{g_W}{2M_W}H(x)\sum_{q=u,d,s}m_q\bar{q}(x)q(x)
\nonumber\\
&=&-\frac{g_W}{4M_W}H(x)\bigg[m_\pi^2\Big(\pi^-(x)\pi^+(x)+\pi^0(x)\pi^0(x)\Big)
         +m_{K^+}^2\, K^+(x)K^-(x)
\nonumber\\&&
+m_{K^0}^2\, K^0(x)\bar{K}^0(x)
+\frac{2\sqrt{2}}{3}\Big(2m_\pi^2-m_{K^+}^2-m_{K^0}^2\Big)
 \eta_1(x)\eta_8(x)
\nonumber\\&&
  +\frac{1}{3}\Big(m_{K^+}^2+m_{K^0}^2+m_\pi^2\Big)\eta_1^2(x) 
  +\frac{1}{3}\Big(2m_{K^+}^2+2m_{K^0}^2-m_\pi^2\Big)\eta_8^2(x)
  \bigg]
\label{HPPL}
\end{eqnarray}
where $H(x)$ is the Higgs field and $\pi^-(x)$, $\pi^+(x)$, $\pi^0(x)$ etc. 
are pseudoscalar-meson fields. Replacing 
the fields $\eta_8(x)$ and 
$\eta_1(x)$ by physical fields $\eta(x)$ and 
$\eta'(x)$ given in Table~II, one obtains the set of 
$H$-boson--pseudoscalar-meson couplings.

The evaluation of the $\langle P_1P_2|\bar{u}(x)\gamma_\mu
d_a(x)\bar{d}_b(y)\gamma_\nu u(y)|0\rangle$ matrix element is, in its full
complexity, a highly nonpertubative problem due to the nonlocality of
the four-quark operators. 
The one-loop pertubative QCD analysis of the $W^+W^-$ diagram shows that
the corresponding amplitude has strong IR divergencies, but no UV
divergencies, even if $W$-propagators are shrunk to points. That suggest
the evaluation of the matrix element in the model which is valid at
very low energies, the gauged $U(3)_L\times U(3)_R/U(3)_V$ chiral model
with pseudoscalar mesons coupled to
the WS gauge bosons. The calculations in the chiral model show that the
contributions to the amplitude come only from the diagrams with pseudoscalar
mesons emitted from different space time points. In the quark picture
that would correspond to splitting of the hadronic matrix element (\ref{SWpWm})
into two vacuum to pseudoscalar-meson matrix elements of the two quark 
operators,
\begin{eqnarray}
\label{hadPP0}
\lefteqn{
\langle P_1 P_2|\bar{u}(x)\gamma_\mu(1-\gamma_5)
 d_a(x)\bar{d}_b(y)\gamma_\nu(1-\gamma_5)u(y)|0\rangle }
\nonumber\\&&
\approx \langle P_1|\bar{u}(x)\gamma_\mu\gamma_5 d_a(x)|0\rangle
 \langle P_2|\bar{d}_b(y)\gamma_\nu\gamma_5 u(y)|0\rangle
+ (P_1\leftrightarrow P_2)
\nonumber\\&&
 =\
2f_{P_1}f_{P_2}\delta_{P_1P(ud_a^c)}\delta_{P_2P(d_bu^c)}
               e^{ip_1x}e^{ip_2y}p_{1\mu}p_{2\nu}+
(P_1\leftrightarrow P_2),\qquad\qquad\qquad
\end{eqnarray}
where $P(ud_a^c)$ and $P(d_bu^c)$ are pseudoscalar mesons having quantum
numbers of the combinations of quarks in brackets ($q^c$ is symbol for
antiquark). Both chiral model approach and quark model approach, in which
(\ref{hadPP0}) is assumed, give the same results. Although the obtained
result is appealing, one must have in mind that chiral models work 
for momentum transfers $\stackrel{\displaystyle <}{\sim}
\ 1\ GeV^2$. Therefore, it is
worth to compare this results with results obtained by some other method,
e.g. sum rules. In the sum rule approach, it is quite unlikely that one can 
split the matrix element as in Eq. (\ref{hadPP0}), and consequently 
the quarks coming from the different space-time points are expected 
to form the (neutral) pseudoscalar mesons, also. That somewhat lessens the 
value of the approximation (\ref{hadPP0}). Unfortunately, the matrix
element with two light pseudoscalar mesons
in the final state cannot be treated by
usual sum-rule techniques as in case of processes with only one light 
pseudoscalar meson in the final state, as for instance in
$D^*\to D\pi$ decays \cite{AKho}, 
because of complications of large
distance strong interactions. The approximation 
(\ref{hadPP0}) will be used
here, because from phenomenology it is known that such approximation 
can hardly fail the correct value of the amplitude by a factor larger
than 5, and because chiral model calculation suggest that
approximation.

Following the procedure outlined above, one obtains the expression for
the generic $T(\tau^-\to l'^-P_1P_2)$ matrix element
\begin{eqnarray}
\label{Ttot}
T(\tau^-\to l'^-P_1P_2)&=& 
\bar{u}_{l'}\gamma_\mu(1-\gamma_5)u_\tau\:
\big(A_{P_1P_2}^{\tau l'}(p_1-p_2)^\mu+B_{P_1P_2}^{\tau l'}q^\mu\big)
\nonumber\\
&&+\bar{u}_{l'}\frac{i\sigma_{\mu\alpha}q^\alpha}{q^2}\:
(m(1+\gamma_5)+m'(1-\gamma_5))u_\tau\,C_{P_1P_2}^{\tau l'}(p_1-p_2)^\mu
\nonumber\\
&&+\bar{u}_{l'}(1+\gamma_5)u_\tau\: D_{P_1P_2}^{\tau l'}
+\bar{u}_{l'}(1-\gamma_5)u_\tau\: E_{P_1P_2}^{\tau l'}
\nonumber\\
&&+\bar{u}_{l'}\not p_2(\not p-\not p_1)\not p_1(1-\gamma_5)u_\tau
F_{P_1P_2}^{\tau l'}
\end{eqnarray}
The first two terms belong to the $\gamma$, $Z$-boson and box amplitude, the
third and fourth to the Higgs-boson amplitude,
and the last one to the $W^+-W^-$ amplitude.
The composite form factors $A_{P_1P_2}^{\tau l'}$, $B_{P_1P_2}^{\tau l'}$,
$C_{P_1P_2}^{\tau l'}$, $D_{P_1P_2}^{\tau l'}$, $E_{P_1P_2}^{\tau l'}$
and $F_{P_1P_2}^{\tau l'}$ read
\begin{eqnarray}
\label{ABCDEF}
A_{P_1P_2}^{\tau l'}&=&
-\sum_{V^0}p_{BW}^{V^0}(q)C_{V^0P_1P_2}\: 
i(a_{V^0}^{\tau l'}+b_{V^0}^{\tau l'})
\nonumber\\
B_{P_1P_2}^{\tau l'}&=&
\sum_{V^0}p_{BW}^{V^0}(q)C_{V^0P_1P_2}\:
i(a_{V^0}^{\tau l'}+b_{V^0}^{\tau l'})\frac{m_1^2-m_2^2}{M_{V_0}^2}
\nonumber\\
C_{P_1P_2}^{\tau l'}&=&
\sum_{V^0}p_{BW}^{V^0}(q)C_{V^0P_1P_2}\: 
i c_{V^0}^{\tau l'}
\nonumber\\
D_{P_1P_2}^{\tau l'}&=&
-\frac{i\alpha_W^2}{16M_W^2}\frac{M^2_{HP_1P_2}}{M^2_H}\: m
\: F_H^{\tau l'}
\nonumber\\
E_{P_1P_2}^{\tau l'}&=&
-\frac{i\alpha_W^2}{16M_W^2}\frac{M_{H^0P_1P_2}^2}{M_{H^0}^2}\: m'
\: G_H^{\tau l'}
\nonumber\\
F_{P_1P_2}^{\tau l'}&=&i\frac{\alpha_W^2\pi^2}{M_W^4}
V_{ud_a}V^\ast_{ud_b}f_{P_1}f_{P_2}F_{W^-W^+}^{\tau l'},
\end{eqnarray}
where
\begin{equation}
p^{V^0}_{BW}=\frac{1}{t-m^2_{V^0}+im_{V^0}\Gamma_{V^0}}
\end{equation}
is a denominator-part of Breit-Wigner propagator for a vector meson
$\tilde{V}^0$ (\ref{BWP}).
$C_{V^0P_1P_2}$ are $V^0-P_1-P_2$ couplings defined by Lagrangian
(\ref{LchSIM}), $a_{V^0}^{\tau l'}$,
$b_{V^0}^{\tau l'}$,and $c_{V^0}^{\tau l'}$ are composite form factors
for $\tau\to l'V^0$ decays found in Ref.~\cite{PRD} and 
listed in Appendix~B,
and $F_{W^-W^+}^{\tau l'}$ is the tree-level form factor,
\begin{equation}
F_{W^-W^+}^{\tau l'} = \frac{1}{(p-p_1)^2}\sum_{N_i}B_{l'N_i}B^\ast_{\tau N_i}.
\end{equation}

Here few comments are in order.
\begin{itemize}
\begin{enumerate}
\item From the structure of the total amplitude (\ref{Ttot}), one can easily
find which of the amplitudes $T_\gamma$, $T_Z$, $T_{Box}$, $T_H$
and $T_{W^-W^+}$ give the dominant contribution. The amplitudes
$T_\gamma$, $T_Z$ and $T_{Box}$ contain a common factor 
$(i\alpha_W^2/16M_W^2)(g_{\rho\pi\pi}/\gamma_V)$.
In place of that factor, in the amplitudes $T_H$ and $T_{W^-W^+}$
are factors $(i\alpha_W^2/16M_W^2)(M^2_{HP_1P_2}/M^2_H)$
and $(i\alpha_W^2\pi^2/M_W^2)
(f_{P_1}f_{P_2}/M_W^2)V_{ud_a}V_{ud_b}^\ast\sum_{N_i}B_{l'N_i}
B^\ast_{\tau N_i}$, respectively.
The amplitudes
$T_Z$ and $T_H$ contain loop form factors behaving as the square of the 
heavy neutrino mass, $m_N^2$, in the large-$m_N$ limit,  
$T_\gamma$ and $T_{Box}$ have $\ln\,m_N$ asymptotics in that limit, and
$T_{W^-W^+}$ is almost independent on $m_N$. Approximating roughly all
momenta of outer particles with $\tau$-lepton mass, one obtains
approximate ratio of the magnitudes of the amplitudes 
\begin{eqnarray}
\label{ratio}
T_{\gamma,Z,Box}\ :\ T_H\ :\ T_{W^-W^+}&\approx&
\frac{g_{\rho\pi\pi}}{\gamma_\rho}
F_Z^{\tau l'}\ :\ \frac{M^2_{HP_1P_2}}{M^2_H}F_H^{\tau l'}
\nonumber\\
&:&16\pi^2\frac{f_{P_1}f_{P_2}}{M_W^2}V_{ud_a}V^\ast_{ud_b} 
\sum_{N_i}B_{l'N_i}B^\ast_{\tau N_i}\qquad
\end{eqnarray}
For heavy-light neutrino mixings $(s_L^{\nu_e})^2=0.01$, 
$(s_L^{\nu_\mu})^2=0$ and ($s_L^{\nu_\tau})^2=0.05$, 
$F_Z^{\tau l'}$ and $F_H^{\tau l'}$ assume values 
$-0.01$ and $0.01$, respectively,
for $m_N=100\: GeV$, and values $-1.6$ and $2.2$, respectively, for 
maximal value of $m_N$ allowed by the pertubative unitarity relation
[see Eq.~(\ref{pertunit}) below], $m_N=3700\: GeV$. Putting these values
into Eq.~(\ref{ratio}), one finds that the $T_{W^-W^+}$ and $T_H$ amplitudes
are six to four and four orders of magnitude smaller than 
the $T_{\gamma,Z,Box}$ amplitude, respectively. The numerical study of relative 
$T_{\gamma,Z,Box}$, $T_H$ and $T_{W^-W^+}$ contributions to the 
$\tau^-\to l'P_1P_2$ branching ratios shows that the $T_{\gamma,Z,Box}$
amplitude participates even more than forseen by this rough estimate.
Therefore, one can safely neglect $H$ and $W^-W^+$ contributions in the
expressions for the largest branching ratios. Since within approximation 
(\ref{hadPP0}) only $T_H$ amplitude participates to 
$\tau^-\to l'^-\pi^0\pi^0/l'^-\eta\eta/l'^-\eta\eta'$ channels, 
it will be kept for illustration of magnitudes of corresponding
branching ratios in Fig.~2.
\item As mentioned in Introduction, the hadronic matrix elements
are evaluated using the non-gauged $U(3)_L\times U(3)_R/U(3)_V$ 
Lagrangian containing hidden 
$U(3)_{local}$ local symmetries. The effective gauge-boson--meson
couplings are introduced through the gauge-boson--quark couplings and 
PCAC (\ref{PCAC}) and vector meson dominance (\ref{EVMD}) relations.
The corresponding effective Lagrangians for vector-boson--$\gamma$ and 
vector-boson--$Z$ interactions read
\begin{eqnarray}   
\label{LVgZ_q} 
{\cal L}_{\gamma V^0}&=&-eA^\mu\Big(
\frac{m_\rho^2}{2\gamma_\rho}\rho_\mu^0+
\frac{m_\phi^2}{2\sqrt{3}\gamma_\phi}c_V\phi_\mu^0+
\frac{m_\omega^2}{2\sqrt{3}\gamma_\omega}s_V\omega_\mu^0\Big)
\nonumber\\
{\cal L}_{Z V^0}&=&-\frac{g_W}{4c_W}Z^\mu
\Big[\frac{m_\rho^2}{\gamma_\rho}c_{2W}\rho_\mu^0+
\frac{m_\phi^2}{\gamma_\phi}
\Big(\frac{c_Vc_{2W}}{\sqrt{3}}+\frac{s_V}{\sqrt{6}}\Big)\phi_\mu
\nonumber\\
&&+\frac{m_\omega^2}{\gamma_\omega}
\Big(\frac{s_Vc_{2W}}{\sqrt{3}}-\frac{c_V}{\sqrt{6}}\Big)\omega_\mu\Big],
\end{eqnarray}
where $s_V = \sin\theta_V$ and $c_V = \cos\theta_V$.
The $\gamma$, $Z$ and $W^-W^+$ amplitudes could be also evaluated using the
gauged version of the $U(3)_L\times U(3)_R/U(3)_V$
chiral Lagrangian with hidden $U(3)_{local}$ symmetry. 
Both approaches give the same results for these
amplitudes. That follows from the comparison of the effective
Lagrangians (\ref{LVgZ_q}) and the corresponding terms in the gauged
chiral Lagrangian (\ref{LchSIM}). Identifying
\begin{equation}
agf_\pi^2\ =\ \frac{m_\rho^2}{2\gamma_\rho}
\ =\ \frac{m_\phi^2}{2\gamma_\phi}
\ =\ \frac{m_\omega^2}{2\gamma_\omega},
\end{equation}
the Lagrangians (\ref{LVgZ_q}) and the coresponding parts of the Lagrangian
(\ref{LchSIM})
become equal. This identification is justified numerically. The 
same type of identification for $W$-boson--pseudoscalar-meson couplings
is trivial, because both approaches use the same hadronic
parameters,
pseudoscalar-meson decay constants.
The indirect way to evaluate hadronic part of the amplitudes
was chosen because the $T_{Box}$ and ${T_H}$ amplitudes do not have their
chiral model counterparts. Moreover, 
this approach enables one to use the experimental
values for the meson masses and branching ratios. In the chiral 
model, they are determined by the symmetries of the model.
\item The chiral nonlinear Lagrangian based on the $U(3)_L\times
U(3)_R/U(3)_V$ symmetry (without hidden symmetries) describes well
the treshold processes \cite{Palmer,BandoPRep} with pseudoscalar 
mesons in the final state only, i.e.  amplitudes of vanishing 
pseudoscalar momenta. To comprise the 
dominant two-pseudoscalar channels of the final state interactions which
swich on at higher energies, vector mesons are introduced
. One of the most common ways to include the
effects of presence of vector mesons into the low energy chiral model
amplitudes is to multiply them with the Breit-Wigner propagators 
normalized to unity at zero-momentum transfer. The constant-width
normalized Breit-Wigner propagator has the following form
\cite{FWWACS,Decker,Palmer},
\begin{equation}
\label{normalisation}
\frac{M_{\tilde{V}}^2-iM_{\tilde{V}}\Gamma_{\tilde{V}}}
{M_{\tilde{V}}^2-t-iM_{\tilde{V}}\Gamma_{\tilde{V}}},
\end{equation}
where $M_{\tilde{V}}$ and $\Gamma_{\tilde{V}}$ 
are vector meson mass and decay width,
respectively. The $\gamma$, $Z$ and box amplitudes obtained in the formalism
of this paper have almost the same structure,
\begin{eqnarray}
\label{chiral}
T_{\gamma,Z,Box}&=&L^\mu_{\gamma,Z,Box}\langle (P_1P_2)_{\tilde{V}^0}
|V_\mu^{\gamma,Z,Box}
-A_\mu^{\gamma,Z,Box}|0\rangle\times K_{\gamma,Z,Box}
\nonumber\\
&&\times\frac{M_{\tilde{V}^0}^2}
{M_{\tilde{V}^0}^2-t-iM_{\tilde{V}^0}\Gamma_{\tilde{V}^0}}
\end{eqnarray}
where $L^\mu_{\gamma,Z,Box}$ are loop parts of the $\tau^-\to l'^-
P_1P_2$ amplitude defined in Eqs.~(\ref{Tg_V}, \ref{TZ_V} and \ref{TBox_V}),
 $K_{\gamma,Z,Box}$ are factors containing coupling
constants ($K_\gamma=-ie$, $K_Z=-ig_W/4c_W$ and $K_{Box}=1$),
and $\langle (P_1P_2)_{\tilde{V^0}}|V_\mu^{\gamma,Z,Box}
-A_\mu^{\gamma,Z,Box}|0\rangle$ comprise products of a 
vacuum-to-vector meson amplitudes of a quark current devided by square
of the vector meson mass, a denominator of the 
vector-meson propagators and a vector-meson--pseudoscalar-meson vertex.
The factor $M_V^2$, which devides the vacuum-to-vector meson amplitude of
the quark current, is extracted from the composite form factors for the 
$\tau^-\to l'^-\tilde{V}^0$, $a_{V^0}^{\tau l'}$, $b_{V^0}^{\tau l'}$
and
$c_{V^0}^{\tau l'}$, and is assigned to the vector meson propagator. 
The low energy limit of the matrix elements 
$\langle P_1P_2|V_\mu^{\gamma,Z,Box}-A_\mu^{\gamma,Z,Box}|0\rangle$
may be derived from the kinetic part of the 
chiral part of the Lagrangian (\ref{LchSIM}), 
$(f_\pi^2/4)\mbox{Tr}(\partial_\mu
U \partial^\mu U^\dagger)$, identifying the quark vector currents 
with the corresponding pseudoscalar-meson vector currents which may be
found in Appendix A.
These low energy limit amplitudes coincide with the corresponding
amplitudes in Eq.~(\ref{chiral}) for zero momentum transfer if the
replacement
\begin{equation}
\label{norm}
M_V^2\to M_V^2-iM_V\Gamma_V
\end{equation}
is made, if 
\begin{equation}
\label{gammas}
\gamma_\rho=\gamma_\omega=\gamma_\phi
\end{equation} 
and if the 
identification 
\begin{equation}
\label{KSFR}
\frac{1}{2\gamma_\rho}\frac{ga}{2}=1
\end{equation}
is made. The equality of the factors $\gamma_{\tilde{V}^0}$ is a consequence
of the $U(3)_L\times U(3)_R/U(3)_V$ 
symmetry, and relation (\ref{KSFR}) is nothing but the
famous Kawarabayashi-Suzuki-Riazuddin-Fayazuddin relation \cite{KSFR}.
Therefore, only the replacement (\ref{norm}) has no natural explanation.
It will be included "by hand", by replacing 
\begin{equation}
\frac{M_{\tilde{V}}^2}{\sqrt{2}\gamma_{\tilde{V}}}\to
\frac{M_{\tilde{V}}^2-iM_{\tilde{V}}\Gamma_{\tilde{V}}}
{\sqrt{2}\gamma_{\tilde{V}}}
\end{equation}
in the vector-meson-dominance relation (\ref{EVMD}).
\item The Lagrangian (\ref{LchSIM}) has $U(3)_L\times U(3)_R/U(3)_V$ symmetry.
The breaking of that symmetry 
will be introduced in the way of Bando, Kugo and Yamawaki 
\cite{BandoNPB} by adding extra terms in the Lagrangian [compare Eqs.~
(\ref{LchSIM}) and (\ref{LchBR})] and by renormalizing the pseudoscalar
fields. In that way, the hidden $U(3)_{local}$ symmetry, which becomes
dependent on $U(3)_L\times U(3)_R$ symmetry through the gauge fixing,
is also broken. Since the Bando et al. Lagrangian is not hermitean,
the Lagrangian in Eq.~(\ref{LchBR}) is written as half of the sum of
their Lagrangian and its hermitean conjugate. Assuming the ideal mixing between 
$SU(3)$-octet and $SU(3)$-singlet vector meson states, 
$\theta_V=\arctan(1/\sqrt{2})$, Bando, Kugo
and Yamawaki obtained the following relations between pseudoscalar 
decay constants, vector meson
masses and vector meson gauge coupling constants 
\begin{eqnarray}
\label{Bando}
f_\pi&=&\frac{f_K}{\sqrt{1+C_A}},
\nonumber\\
m_\rho^2&=&m_\omega^2\ =\ ag^2f_\pi^2\ =\ \frac{m_{K^\ast}^2}{1+C_V}
\ =\ \frac{m_\phi^2}{(1+C_V)^2}, 
\nonumber\\
\frac{g_{\gamma\rho}}{m_\rho^2}&=&\frac{3g_{\gamma\omega}}{m_\omega^2}\ =\ 
-\frac{3g_{\gamma\phi}}{\sqrt{2}m_\phi^2}\ =\ \frac{1}{g},
\end{eqnarray}
where $C_A$ and $C_V$ are breaking parameters appearing in the Lagrangian
(\ref{LchBR}), and 
$g_{\gamma\rho}$, $g_{\gamma\omega}$ and $g_{\gamma\phi}$ are
gauge-boson--vector meson coupling constants which may be 
read from the Lagrangians (\ref{LchSIM}) and (\ref{LchBR}). 
Replacing the expressions for the gauge coupling constants
from Eq.~(\ref{Bando}) with the corresponding expressions in the 
Lagrangians (\ref{LVgZ_q}) into the third of Eqs.~(\ref{Bando}), one
obtains again Eq.~(\ref{gammas}). Therefore, if the ideal mixing  
between $SU(3)$-octet and $SU(3)$-singlet vector mesons is 
assumed, the equality of $\gamma_{\tilde{V}^0}$-s is preserved after
the symmetry breaking. In this paper, the ideal mixing condition 
is relaxed:
the mixing angle $\theta_V$ is evaluated from the experimental meson masses
using the quadratic Gell-Mann--Okubo mass formula.
\end{enumerate}
\end{itemize}

Keeping in mind the above comments, one can derive the corresponding
expression for the branching ratios from
the expression for the generic $\tau^-\to l'^-P_1P_2$ amplitude: 
\begin{eqnarray}
B(\tau^-\to l'^-P_1P_2)&=&\frac{1}{256\pi^3m^3\Gamma_\tau}
\int_{(m_1+m_2)^2}^{(m-m')^2}\; dt\int_{s_1^-}^{s_1^+} ds_1
\langle |T(\tau^-\to l'^-P_1P_2)|^2\rangle
\nonumber\\
&=&\frac{1}{64\pi^3m^3\Gamma_\tau}\int_{(m_1+m_2)^2}^{(m-m')^2}dt
\Big[\alpha |A_{P_1P_2}^{\tau l'}|^2
+\beta (A_{P_1P_2}^{\tau l'}B_{P_1P_2}^{\tau l'*}+h.c.)
\nonumber\\&&
+\gamma |B_{P_1P_2}^{\tau l'}|^2
-\delta (A_{P_1P_2}^{\tau l'}C_{P_1P_2}^{\tau l'*}+h.c.)
-\varepsilon |C_{P_1P_2}^{\tau l'}|^2
\nonumber\\&&
+\zeta \Big(A_{P_1P_2}^{\tau l'}(D_{P_1P_2}^{\tau l'*}
  +\frac{m'}{m}E_{P_1P_2}^{\tau l'*})+h.c.\Big)
\nonumber\\&&
+\eta \Big(B_{P_1P_2}^{\tau l'}(D_{P_1P_2}^{\tau l'*}
  +\frac{m'}{m}E_{P_1P_2}^{\tau l'*})+h.c.\Big)
\nonumber\\&&
+\vartheta \Big(C_{P_1P_2}^{\tau l'}(D_{P_1P_2}^{\tau l'*}
  +\frac{m'}{m}E_{P_1P_2}^{\tau l'*})+h.c.\Big)
\nonumber\\&&
+\iota (|D_{P_1P_2}^{\tau l'}|^2
  +|E_{P_1P_2}^{\tau l'}|^2)
+\kappa(D_{P_1P_2}^{\tau l'}E_{P_1P_2}^{\tau l'*}+h.c.)\Big],
\end{eqnarray}
where integration boundaries $s_1^\pm$ 
and parts of the square of the amplitude depending on the momentum
transfer variable $t$,\  $\alpha$, $\beta$, $\gamma$, $\delta$, 
$\varepsilon$, $\zeta$, $\eta$,
$\vartheta$, $\iota$ and $\kappa$ may be found in Appendix C.

\section{{Numerical results}}
\setcounter{equation}{0}

In the numerical analysis, the extension of the SM with two heavy neutrinos is
assumed. The description of the model and the relevant formulas for $B$ and $C$
matrices may be found in Introduction. The additional parameters 
of the model
are three heavy-light mixings, $s_L^{\nu_l}$, and two heavy neutrino masses, 
$m_{N_1}$ and $m_{N_2}$. The upper limits (\ref{upli}) and (\ref{upli2})
experimentally constrain the mixings $s_L^{\nu_l}$, while the upper 
bound on heavy neutrino masses,
\begin{equation}
\label{pertunit}
m_{N_1}^2\leq\frac{2M_W^2}{\alpha_W}\frac{1+\rho^{-1/2}}
{\rho^{1/2}}\Big[\sum_i(s_L^{\nu_i})^2\Big]^{-1},
\qquad \rho\ \geq\ 1
\end{equation}
may be obtained from the perturbative unitarity relations
\cite{NPB,AP_pertu,CFH}. 
The experimental upper bound limits (\ref{upli}) suggest that either 
$s_L^{\nu_e}$
or $s_L^{\nu_\mu}$ is approximately equal zero. Here will be assumed that 
$s_L^{\nu_\mu}\approx 0$, and, therefore, only 
$\tau^-\to e^\mp P_1P_2$ decays are
considered. The results obtained for $s_L^{\nu_e}\approx 0$ case,
that is for $\tau^-\to \mu^\mp P_1P_2$ decays, almost coincide with 
corresponding $s_L^{\nu_\mu}\approx 0$ results, and it is superfluous to 
discuss them seperately. The $\tau^-\to e^\mp P_1P_2$ decays depend on
new parameters of the model, $s_L^{\nu_e}$, $s_L^{\nu_\tau}$ and
$m_{N_i}$, as well as on a whole set of quark-level parameters and meson
observables: CKM mixing angles, quark masses, mixing angle between octet
and singlet vector-meson states, meson masses and decay widths,
pseudoscalar-meson decay constants, constants describing the coupling
strength of vector mesons to the gauge bosons and
vector-meson--psudoscalar-meson coupling constants. In calculations, the
average of the experimental upper and lower values for CKM matrix elements
are used, and the quark masses
\begin{eqnarray}
m_u&=&0.005~\mbox{GeV},\qquad m_d\ =\ 0.010\mbox{GeV},\qquad 
m_s\ =\ 0.199~\mbox{GeV},
\nonumber\\
m_c&=&1.35~\mbox{GeV},\qquad m_b\ =\ 4.3~\mbox{GeV},\qquad 
m_t\ =\ 176~\mbox{GeV},
\end{eqnarray}
cited in Refs.~\cite{PDG94,qm}. The masses off all quarks are kept in
evaluation of matrix elements, since $t$ and $c$ quarks give comparable
contributions to some amplitudes. The mixing angle between singlet and
octet vector-meson states is not taken to be equal to the ideal-mixing
value, $\theta_V=\arctan(1/\sqrt{2})$, but is either determined
from the quadratic Gell-Mann--Okubo mass formula, or treated as a free
parameter. For pseudoscalar decay constants $f_{\pi^\pm}$ and
$f_{K^\pm}$, appearing only in the $W^+W^-$ amplitudes of 
$\tau^-\to e^+P_1^-P_2^-$ decays and $\tau^-\to e^+P_1^-P_2^-$
amplitudes, the experimental values are used \cite{PDG94}
\begin{equation}
f_{\pi^\pm}\ =\ 92.4\ MeV,\qquad f_{K^\pm}\ =\ 113\ MeV.
\end{equation}
The constants $\gamma_{\tilde{V}}$, describing the coupling strengths of
vector mesons to the gauge bosons, are either extracted from
$\tilde{V}\to e^+e^-$ decay rates
\begin{equation}
\gamma_{\rho^0}\ =\ 2.519,\qquad \gamma_\omega\ =\ 2.841,\qquad
\gamma_\phi\ =\ 3.037,
\end{equation}
or estimated using $SU(3)$-octet symmetry:
$\gamma_{K^{0\ast}}=\gamma_{\rho^0}$. Notice that the equality of 
$\gamma_{\tilde{V}^0}$-s predicted by $U(3)_L\times U(3)_R/U(3)_V$
symmetric chiral model and by $U(3)_L\times U(3)_R/U(3)_V$ broken 
chiral model is reasonably satisfied. The decay rates of vector mesons,
involved through the vector-meson propagators, are taken to be equal to
their experimental total-decay-rate values \cite{PDG94}, and are not
treated as momentum dependent quantities \cite{Decker}. The 
$\rho$-$\pi$-$\pi$ coupling is derived from  the $\rho\to 2\pi$ decay
width, while the other vector-meson--pseudoscalar-meson couplings are
fixed by one of the chiral models described in Appendix A. It is
visible from the above that, whenever possible, the parameters were
extracted from experiment and model dependent relations determining them
were relaxed.

In this paper, 17 \ $\tau^-\to e^\mp P_1P_2$ decays are studied numerically.
For orientation of the reader, decay widths of all 17 reactions are
plotted in Fig.~2 as functions of $m_{N_1}=\frac{1}{3}m_{N_2}$ for 
upper bound values of heavy-light neutrino mixings (\ref{upli}).
Concerning the $m_{N_1}$ dependence, the decays can be split into four
groups: $\tau^-\to e^-\pi^+\pi^-/e^-K^+K^-/e^-K^0\bar{K}^0$,
$\tau^-\to e^-\pi^+K^-/e^-\pi^-K^+/e^-\pi^0K^0/e^-\pi^0\bar{K}^0
/e^-K^0\eta/e^-\bar{K}^0\eta/e^-K^0\eta'/e^-\bar{K}^0\eta'$,
$\tau^-\to e^-\pi^0\pi^0/e^-\eta\eta/e^-\eta\eta'$ and $\tau^-\to
e^+\pi^-\pi^-/e^+\pi^-K^-/e^+K^-K^-$. Only the decays of the first group
are interesting from the experimental point of view and receive
contributions from all five $\tau^-\to e^- P_1P_2$ amplitudes [see Eq.~
(\ref{Stot})]. The others are suppressed by at least 8 orders of magnitude
relative to the first group of decays for various reasons. The members
of the second group are Cabbibo suppressed, and only box and $W^+W^-$ 
diagrams contribute to them. The decays of the third group originate
from the $H$-amplitude and are suppressed by the factor 
$(M^2_{HP_1P_2}/M_H^2)^2$ from Eq.~(\ref{ratio}). The last group belongs
to the Majorana-type decays, receives contributions only from tree-level 
amplitudes and is suppressed by two factors: by the factor 
$\sim (T_{W^-W^+}/T_{\gamma,Z,Box})^2$ from Eq.~(\ref{Stot}), and by the
additional factor $\sim(m_\tau^2/m_{N_1}^2)^2$ comming from the heavy
neutrino propagators. In Fig.~2, the choice $m_{N_1}=m_{N_2}/3$
was made since Majorana-type decays vanish if the masses of heavy
neutrinos are equal.

In the following, only the first group of decays is discussed. The results
are given in Figs.~3-6. Figures~3 and~4 show the dependence of the
branching ratios $B(\tau^-\to e^-\pi^+\pi^-/e^-K^+K^-/e^-K^0\bar{K}^0)$
on new weak interaction parameters of the model, $s_L^{\nu_i}$ and
$m_{N_i}$. Figures~5 and~6 illustrate the dependence of these branching
ratios on model assumptions for hadronic part of the amplitude and on 
some strong interaction parameters. 

The Figures~3(a) and~3(b) illustrate $m_N=m_{N_1}=m_{N_2}$ dependence of the
branching ratios for $(s_L^{\nu_e})^2=0.01$ and two different values of 
$(s_L^{\nu_\tau})^2$. The maximum values for branching ratios are
obtained for maximal $m_N$, $(s_L^{\nu_e})^2$ and $(s_L^{\nu_\tau})^2$ 
values permitted by Eqs.~(\ref{pertunit}) and (\ref{upli}):
\begin{eqnarray}
\label{upliBR_th}
B(\tau^-\to e^-\pi^+\pi^-)&\stackrel{\displaystyle <}{\sim}&
0.74\cdot 10^{-6}\ (0.35\cdot 10^{-6})
\nonumber\\
B(\tau^-\to e^-K^+K^-)&\stackrel{\displaystyle <}{\sim}&
0.42\cdot 10^{-6}\ (0.20\cdot 10^{-6})
\nonumber\\
B(\tau^-\to e^-K^0\bar{K}^0)&\stackrel{\displaystyle <}{\sim}&
0.26\cdot 10^{-6}\ (0.12\cdot 10^{-6})
\end{eqnarray} 
The expressions in the parentheses are obtained for the upper bound
$(s_L^{\nu_e})^2$ and $(s_L^{\nu_\tau})^2$ values refered in Eq.~
(\ref{upli2}). The present experimental bound exists only for one of
these decays
\begin{equation}
B(\tau^-\to e^-\pi^+\pi^-)< 4.4\cdot 10^{-6},
\end{equation}
because the main $\tau^-\to\tilde{V}^0$ contribution mode to the 
$\tau^-\to e^-K^+K^-/e^-K^0\bar{K}^0$ decays, $\tau^-\to e^-\phi$, has not
been experimentaly searched for yet.
In Figures~3(a) and~3(b), the branching fractions 
$B(\tau^-\to e^-\pi^+\pi^-/e^-K^+K^-/e^-K^0\bar{K}^0)$ 
are shown. The behaviour of the branching ratio terms quadratic 
and quartic in $s_L^{\nu_i}$ expansion have similar behaviour as the 
corresponding terms in $\tau\to e^-M^0$ decays \cite{NPB,PRD}.
For $m_N$ values below 200 GeV,
quadratic $(s_L^{\nu_i})^2$ terms, that have $\ln (m_M^2/m_W^2)$
large-$m_N$ behaviour, prevail, while for larger $m_N$ quartic terms
having $m_N^2$ large-$m_N$ asymptotics dominate.
As $(s_L^{\nu_\tau})^2$ decreases, the branching fractions also decrease,
but at the same time the pertubative unitarity upper bound on $m_N$
increases, and, therefore, branching ratios increase in the larger $m_N$
interval. These two opposite effects lead to the small difference of the largest
values for branching fractions in Eq.~(\ref{upliBR_th}). The 
nondecoupling behaviour of the
branching ratios displayed in Figs.~3(a) and~3(b) is a consequence of the
implicit assumption that the mixings  $s_L^{\nu_i}$ may be kept
constant in the whole $m_N$-interval of interest. As mentioned in
Introduction, $s_L^{\nu_i}\propto m_D/m_M\propto m_D/m_{N_i}$, and,
therefore, the constancy of $s_L^{\nu_i}$ implies that for large $m_{N_i}$
values, the Dirac components, $m_D$, are large also. 
Since the Dirac-mass values are bounded by the typical SM
$SU(2)\times U(1)$ breaking scale, $v\sim 250$ GeV (more precisely,
pertubative unitarity upper bound on the Dirac mass is $m_D\leq 1\ TeV$ 
\cite{CFH}), this condition
cannot be satisfied in the $m_N\to\infty$ limit, leading to vanishing
effects of heavy neutrinos
\cite{SenjSok}. Nevertheless, for $0.1$ TeV$\leq m_N
\leq 10$ TeV it can be fullfilled. Nondecoupling effects of the heavy 
neutrinos were first studied in Ref. \cite{AP_H}, and were also extensively
studied in Refs. \cite{NPB,PRD,AP_pertu,AP}.

Figures~3(c) and~3(d) present the dependence of the branching ratios on 
$(s_L^{\nu_\tau})^2$ and $(s_L^{\nu_e})^2$ respectively, for $m_N=4000$ GeV.
The branching ratios are almost quadratic functions of
$(s_L^{\nu_\tau})^2$, and almost linear functions of $(s_L^{\nu_e})^2$.
Such dependence is expected from the large-$m_N$ behaviour of form
factors \cite{NPB} (see also Appendix B).

Figure~4 illustrates Majorana-neutrino quantum effects. It displays the
dependence of branching fractions on the ratio $m_{N_2}/m_{N_1}$ for 
fixed values $m_{N_1}=1$ TeV and $m_{N_1}=0.5$ TeV. The maximal
$B(\tau^-\to e^-\pi^+\pi^-/e^-K^+K^-/e^-K^0\bar{K}^0)$ values are
obtained for $m_{N_2}/m_{N_1}\sim 3$. These effects are also a consequence
of large $s_L^{\nu_i}$ mixings (large Dirac components of the neutrino
mass matrix), since they enter through the loop functions depending on
two heavy neutrino masses, which can be found only in quartic 
terms in the $s_L^{\nu_i}$ expansion. A similar behaviour has been
found for $\tau^-\to l'^-M^0$ \cite{PRD} and $\tau^-\to l'^-l_1^-l_2^+$ 
\cite{NPB} decays.

Figures~5(a)-5(c) show the influence of the main ingredients of the hadronic
part of the amplitudes discussed in the comments of Section 2 on the 
branching ratios. Thick lines in Figs.~5(a)-5(c) correspond to the situation
when one of the theoretical assumptions is changed. Thin lines serve as
reference results and they coincide with the complete-calculation graphs
shown in Fig.~3(a).

Figure~5(a) shows the dependence of the branching ratios on vector-meson
resonances. When the vector-meson propagators are replaced by their 
zero-momentum-transfer values, that is when the normalized vector-meson
propagators (\ref{normalisation}) are replaced by 1, one obtains the
chiral-limit values for the branching ratios plotted in Fig.~5(a), which
are considerably smaller. The $B(\tau^-\to e^-\pi^+\pi^-/e^-K^-K^+$) 
branching ratios decrease by factors $\sim 5$ and $\sim 20$ respectively.
The decrease of $\tau^-\to e^-K^-K^+$ branching ratio is more prominent,
because it receives main contribution from the narrower $\phi$
resonance, while to $B(\tau^-\to e^-\pi^+\pi^-)$ only $\rho$-resonance
contributes.
The $\tau^-\to e^-K^0\bar{K}^0$ branching ratio becomes almost equal to zero
because its amplitude is proportional to the expression $F_{Box}^{\tau
l'dd}-F_{Box}^{\tau l'ss}$ which is almost equal to zero.

In Figure~5(b), the $U(3)\times U(3)_R/U(3)_V$ breaking effects are emphasized
by comparing the branching ratios obtained in the $U(3)\times
U(3)_R/U(3)_V$ symmetric chiral model with reference results which
include $U(3)\times U(3)_R/U(3)_V$ symmetry breakings. The symmetry
breaking does not influence $B(\tau^-\to e^-\pi^-\pi^+)$, but 
$B(\tau^-\to e^-K^-K^+/e^-K^0\bar{K}^0)$ are enlarged by a factor $\sim 1.5$.

The reference results include the $\theta_V$-value derived from the
Gell-Mann--Okubo quadratic mass formula, $\theta_V=39.1^\circ$. In Fig.~5(c),
these results are compared with branching ratios evaluated for 
$\theta_V=30^\circ$. As $\theta_V$ is known to be close to the ideal
mixing value $\arctan(1/\sqrt{2})$, the weak $\theta_V$-dependence
displayed in Fig.~5(c) implies that $\theta_V$ variation cannot influence
the branching ratios strongly.

The influence of the replacement (\ref{norm}) induces so small changes
of the branching ratios that they cannot be observed in a figure. For
that reason these results have not been plotted.

The Figure~6 gives the dependence of the partial decay rates on the
momentum transfer variable, $t=(p-p')^2$. 
The $\tau^-\to e^-\pi^-\pi^+$ decay rate
receives the  contribution from the broad $\rho^0$-resonance only.
The $\tau^-\to e^-K^-K^+/e^-K^0\bar{K}^0$ decays receive contributions
from all three flavour-neutral resonances, but for the kinematical
reasons only very narrow $\phi$-resonance can be noticed in the spectra.

\section{{Conclusions}}
\setcounter{equation}{0}

This paper completes the analysis of the experimentally investigated
neutrinoless $\tau$-lepton decays within heavy-Majorana/Dirac-neutrino
extensions of the SM, started in the previous publications \cite{NPB,PRD}.
For the experimentally most promising decays, $\tau^-\to e^-\pi^+K^-/
\mu^-\pi^-K^+/\mu^+\pi^-K^-$, the calculated branching ratios were found
to be much smaller than the current experimental upper bounds.
Nevertheless, the three of seventeen explored decays, $\tau^-\to
e^-\pi^+\pi^-/e^-K^+K^-/e^-K^0\bar{K}^0$, were found to have branching
fractions of the order of $10^{-6}$, and the 
first of them the branching fraction
close to the current experimental sensitivity. The other two decays have
not been measured yet, because the reaction $\tau^-\to e^-\phi$, giving
the main contribution to these decays, has not been experimentally
investigated yet.

The main feature of the leptonic sector of the model used here is largeness
of the heavy-light neutrino mixings $s_L^{\nu_i}$. From it the dominance
of the quartic $s_L^{\nu_i}$ terms and the $m_{N_i}^2$ behaviour of 
$\tau^-\to e^-\pi^+\pi^-/e^-K^+K^-/e^-K^0\bar{K}^0$ in the large-$m_N$  
limit follows, giving rise to the enhancement of the branching ratios by
the factor ~40 relative to the results obtained by the analysis in which the
respective terms are omitted. The $s_L^{\nu_i}$ behaviour and the 
$m_{N_2}/m_{N_1}$ dependence of the branching ratios are also
consequences of large $s_L^{\nu_i}$ mixings. Particularly, the
$m_{N_2}/m_{N_1}$ dependence leads to the maxima of branching ratios for
$m_{N_2}/m_{N_1}\sim 3$, the same as in $\tau^-\to l'^-l_1^-l_2^+$
\cite{NPB} and $\tau^-\to l'^-M^0$ \cite{PRD} decays.

Several ingredients of the hadronic part of the $\tau^-\to l'^-P_1P_2$
amplitudes, that influence the magnitude of the corresponding branching
ratios, were discussed. The most prominent contribution comes from the
vector-meson resonances, giving rise to enhancemens of $B(\tau^-\to
e^-\pi^+\pi^-/e^-K^+K^-)$ by factors $\sim 5$ and $\sim 20$ and making
$B(\tau^-\to e^-K^0\bar{K}^0)$ different from its chiral 
limit value, zero, and approximately equal to branching values of the other
two decays. The narrower resonances lead to larger
enhancements. The $U(3)_L\times U(3)_R/U(3)_V$ breaking of the chiral
symmetry induce smaller changes of the branching ratios, and they
influence only the $\tau^-\to e^-K^+K^-/e^-K^0\bar{K}^0$ branching
fractions. All other modifications of or changes in the hadronic part of
the $\tau^-\to l'^\mp P_1P_2$ amplitudes discussed here have negligible
influence on the branching ratios.

\vspace{2cm}
\noindent
{\bf Acknowledgements.}
I wish to thank B. Kniehl and A. Pilaftsis for useful comments on
pertubative part of lepton violating amplitudes, and to A. Pilaftsis
for carefully reading the manuscript. I am indebted to S. Fajfer for
many discussions, comments and several ideas concerning the chiral
Lagrangians, and the Higgs-exchange amplitude, and for carefully reading the
manuscript, as well as to A.
Khodjamirian for very detailed discussion on the sum rule aspect on
matrix element (\ref{hadPP0}). I am also indebted to the Theory Group of the
Max-Planck-Institut f\"ur Physik for the kind hospitality 
extended to me during a
visit, when part of this work was performed. This work is supported by
the Forschungszentrum J\"ulich GmbH, Germany, under the project number
6B0A1A, and by project 1-03-233 "Field theory and structure of elementary 
particles".
\setcounter{section}{0}
\def\theequation{\Alph{section}.\arabic{equation}}
\begin{appendix}
\setcounter{equation}{0}
\section{Strong interaction Lagrangians}
\indent

The gauged chiral $U(3)_L\times U(3)_R/U(3)_V$ Lagrangian extended by
hidden $U(3)_{local}$ symmetry and the mass term for pseudoscalar mesons
reads
\begin{eqnarray}
\label{LchSIM}
{\cal L}&=&{\cal L}_A+a{\cal L}_V+{\cal L}_{mass}+{\cal L}_{kin}
\nonumber\\
&=& -\frac{1}{4}f_\pi^2
Tr(D_\mu\xi_L\xi_L^\dagger-D_\mu\xi_R\xi_R^\dagger)^2
-\frac{a}{4}f_\pi^2
Tr(D_\mu\xi_L\xi_L^\dagger+D_\mu\xi_R\xi_R^\dagger)^2
+{\cal L}_{mass}+{\cal L}_{kin}
\nonumber\\
&=&\Big\{\frac{f_\pi^2}{4}Tr(\partial_\mu U \partial^\mu U^\dagger)\Big\}
+\bigg\{\frac{f_\pi^2}{4}r\: Tr\Big(m(U+U^\dagger)\Big)\bigg\}
\nonumber\\&&
+\bigg\{-e(agf_\pi^2)\Big(\rho_\mu^0+\frac{c_V}{\sqrt{3}}\phi_\mu
                +\frac{s_V}{\sqrt{3}}\omega_\mu\Big)A^\mu
\nonumber\\&&
-e(agf_\pi^2)\bigg(\frac{1-2s_W^2}{2s_Wc_W}\rho_\mu^0
    +\Big(\frac{c_V}{\sqrt{3}}\frac{1-2s_W^2}{2s_Wc_W}
         +\frac{s_V}{\sqrt{6}}\frac{1}{2s_Wc_W}\Big)\phi_\mu
\nonumber\\&&
    +\Big(\frac{s_V}{\sqrt{3}}\frac{1-2s_W^2}{2s_Wc_W}
         -\frac{c_V}{\sqrt{6}}\frac{1}{2s_Wc_W}\Big)\omega_\mu
\bigg)Z^\mu\bigg\}        
\nonumber\\&&
+\frac{-iga}{4}
\Big\{\rho^{0,\mu}\big(
2\pi^+\hspace{-4pt}\stackrel{\leftrightarrow}{\partial}_\mu\hspace{-2pt}\pi^-
+K^+\hspace{-4pt}\stackrel{\leftrightarrow}{\partial}_\mu
     \hspace{-2pt}K^-
-K^0\hspace{-4pt}\stackrel{\leftrightarrow}{\partial}_\mu
     \hspace{-2pt}\bar{K}^0\big)
\nonumber\\&&
+\sqrt{3}s_V\omega^\mu\big(
  K^+\hspace{-4pt}\stackrel{\leftrightarrow}{\partial}_\mu
     \hspace{-2pt}K^-
 +K^0\hspace{-4pt}\stackrel{\leftrightarrow}{\partial}_\mu
     \hspace{-2pt}\bar{K}^0\big)
+\sqrt{3}c_V\phi^\mu\big(
  K^+\hspace{-4pt}\stackrel{\leftrightarrow}{\partial}_\mu
     \hspace{-2pt}K^-
 +K^0\hspace{-4pt}\stackrel{\leftrightarrow}{\partial}_\mu
     \hspace{-2pt}\bar{K}^0\big)
\nonumber\\&&
+K^{0*,\mu}\Big(
-\sqrt{2}
  \pi^+\hspace{-4pt}\stackrel{\leftrightarrow}{\partial}_\mu
     \hspace{-2pt}K^-
 +\pi^0\hspace{-4pt}\stackrel{\leftrightarrow}{\partial}_\mu
     \hspace{-2pt}\bar{K}^0
 +\sqrt{3}c_P
  \bar{K}^0\hspace{-4pt}\stackrel{\leftrightarrow}{\partial}_\mu
     \hspace{-2pt}\eta
 +\sqrt{3}s_P
  \bar{K}^0\hspace{-4pt}\stackrel{\leftrightarrow}{\partial}_\mu
     \hspace{-2pt}\eta'\Big)
\nonumber\\&&
+\bar{K}^{0*,\mu}\Big(
  \sqrt{2}
  \pi^-\hspace{-4pt}\stackrel{\leftrightarrow}{\partial}_\mu
     \hspace{-2pt}K^+
 -\pi^0\hspace{-4pt}\stackrel{\leftrightarrow}{\partial}_\mu
     \hspace{-2pt}K^0
 -\sqrt{3}c_P
  K^0\hspace{-4pt}\stackrel{\leftrightarrow}{\partial}_\mu
     \hspace{-2pt}\eta
 -\sqrt{3}s_P
  K^0\hspace{-4pt}\stackrel{\leftrightarrow}{\partial}_\mu
     \hspace{-2pt}\eta'\Big)
\Big\}+\dots\qquad
\end{eqnarray}
where ${\cal L}_{kin}$ is the kinetic Lagrangian of gauge fields,
$f_\pi$ is the pseudoscalar decay constant, $a$ is a free
parameter equal $2$ if vector-meson dominance is satisfied, $g$ is the
coupling of (hidden symmetry induced) vector mesons $V$, 
to the chiral fields $\xi_{L,R}$, $c_W=\cos\theta_W$,
\begin{eqnarray}
D_\mu\xi_L(x)&=&(\partial_\mu-iV_\mu(x))\xi_L(x)
+i\xi_L(x){\cal L}_\mu(x), \qquad
\big(L\leftrightarrow R,\; {\cal L}_\mu \leftrightarrow {\cal R}_\mu\big),
\qquad
\\
\xi_{L,R}(x)&=&e^{i\sigma(x)/f_\pi}
e^{\mp i\pi(x)/f_\pi},\qquad \sigma(x)=0,
\end{eqnarray}
$\sigma(x)=0$ beeing special (unitary) gauge choice. ${\cal L}_\mu(x)$ and
${\cal R}_\mu(x)$ are combinations of gauge fields,
\begin{eqnarray}
{\cal L}_\mu(x)&=&eQ\Big(A_\mu(x)-t_WZ_\mu(x)\Big)
 +\frac{e}{s_Wc_W}T_zZ_\mu(x)+\frac{e}{\sqrt{2}s_W}W_\mu,
\nonumber\\
{\cal R}_\mu(x)&=&eQ\Big(A_\mu(x)-t_WZ_\mu(x)\Big),
\end{eqnarray}
where
\begin{eqnarray}
Q&=&\frac{1}{3}
\left(\begin{array}{ccc}
2&0&0\\
0&-1&0\\
0&0&-1
\end{array}\right),\quad
T_z=\frac{1}{2}
\left(\begin{array}{ccc}
1&0&0\\
0&-1&0\\
0&0&-1
\end{array}\right),
\end{eqnarray}
are quark charge and isospin matrices, 
\begin{eqnarray}
W_\mu(x)&=&
\left(\begin{array}{ccc}
0&W_\mu^+(x)c_c&W_\mu^+(x)s_c\\
W_\mu^-(x)c_c&0&0\\
W_\mu^-(x)s_c&0&0
\end{array}\right),\qquad
\end{eqnarray}
$c_c$ and $s_c$ being cosine and sine of the Cabbibo angle, respectively.
$A_\mu(x)$, $Z_\mu(x)$ and
$W_\mu^\pm(x)$ are photon, $Z$-boson and $W_\mu^\pm$-boson fields.
The dots in Eq.~(\ref{LchSIM}) represent remaining terms in the 
gauged chiral $U(3)_L\times U(3)_R/U(3)_V$ Lagrangian containing 
hidden $U(3)_{local}$ symmetry, not interesting for the topics discussed
in this paper. The first curly bracket contains minimal non-gauged 
chiral model Lagrangian. Using the Gell-Mann--L\'evy procedure \cite{GML},
the pseudoscalar-meson vector currents 
may be derived from that Lagrangian: 
\begin{equation}
V_\mu^a(x)=-2Tr\Big\{T^a[\pi(x),\partial_\mu\pi(x)]\Big\},
\end{equation}
with $\pi(x)$ and $T^a$ defined below Eq.~(\ref{q_scalar}). For instance
the vector current having quantum numbers of $\rho$ meson reads
\begin{equation}
\frac{1}{\sqrt{2}}V_\mu^3=
\frac{1}{\sqrt{2}}\pi^+\hspace{-4pt}\stackrel{\leftrightarrow}
   {\partial}_\mu\hspace{-2pt}\pi^-
+\frac{1}{2\sqrt{2}}K^+\hspace{-4pt}\stackrel{\leftrightarrow}
   {\partial}_\mu\hspace{-2pt}K^-
-\frac{1}{2\sqrt{2}}K^0\hspace{-4pt}\stackrel{\leftrightarrow}
   {\partial}_\mu\hspace{-2pt}\bar{K}^0
\end{equation}
Pseudoscalar mass terms may be found in the second curly bracket.
The $m$ is a mass matrix of $u$, $d$ and $s$ quarks, and $r$ is defined
in Eq. (\ref{r}).
Terms in the third curly bracket represent
photon--vector-boson and Z-boson--vector-boson interactions. These
interactions define the corresponding gauge-boson--vector-meson coupling
strengths (for instance photon--$\rho$-meson couping is equal to 
$-eagf_\pi^2$). The fourth curly bracket comprises
vector-meson--two-pseudoscalar-meson interactions and defines the
corresponding couplings.

The breaking of the $U(3)_L\times U(3)_R/U(3)_V$ symmetry is introduced
in the way of Bando, Kugo and Yamawaki \cite{BandoNPB}. Besides the
terms containing only the $\xi_L$ or $\xi_R$ fields, they added the
additional mixing terms, combined with the matrix-valued parameters,
\begin{equation}
\varepsilon_{A,V}=
\left(\begin{array}{ccc}
0&&\\
&0&\\
&&C_{A,V}
\end{array}\right)
\end{equation}
defining the magnitude of the symmetry breaking. These additional terms
change the kinetic part of the pseudoscalar-field Lagrangian. To restore
the original form of kinetic terms pseudoscalar-meson 
 fields have to be renormalized:
\begin{equation}
\pi(x)\to \pi^r(x)\equiv \lambda_A^{1/2}\pi(x)\lambda_A^{1/2}
\end{equation}
where $\lambda_{A,V}=1+\varepsilon_{A,V}$. 
Following the described procedure, 
one obtains the following expression 
\begin{eqnarray}
\label{LchBR} 
{\cal L}^{br}&=&{\cal L}_A^{br}+a{\cal L}_V^{br}+{\cal L}_{mass}+{\cal
L}_{kin}
\nonumber\\
&=&\Big[\Big\{-\frac{1}{8}f_\pi^2
Tr\big((D_\mu\xi_L\xi_L^\dagger+D_\mu\xi_L\varepsilon_A\xi_R^\dagger)
  -(D_\mu\xi_R\xi_R^\dagger+D_\mu\xi_R\varepsilon_A\xi_L^\dagger)\big)^2
\nonumber\\&&
-\frac{a}{8}f_\pi^2
Tr\big((D_\mu\xi_L\xi_L^\dagger+D_\mu\xi_L\varepsilon_V\xi_R^\dagger)
  +(D_\mu\xi_R\xi_R^\dagger+D_\mu\xi_R\varepsilon_V\xi_L^\dagger)\big)^2
\Big\}+{h.c.}\Big]
\nonumber\\&&
+{\cal L}_{mass}+{\cal L}_{kin}
\nonumber\\
&=&\Bigg\{-e(agf_\pi^2)\bigg(\rho_\mu^0
+\Big(\frac{1+2\gamma_V^2}{3\sqrt{3}}c_V
 -\frac{2(1-\gamma_V^2)}{3\sqrt{6}}s_V\Big)\phi_\mu
+\Big(\frac{1+2\gamma_V^2}{3\sqrt{3}}s_V
\nonumber\\&&
 -\frac{2(1-\gamma_V^2)}{3\sqrt{6}}c_V\Big)\omega_\mu\bigg)A^\mu
-e(agf_\pi^2)\Bigg(\frac{1-2s_W^2}{2s_Wc_W}\rho_\mu^0
\nonumber\\&&
 +\bigg(\frac{c_V}{\sqrt{3}}\frac{1}{2s_Wc_W}
   \Big(\gamma_V^2-2s_W^2\frac{1+\gamma_V^2}{3}\Big)
 +\frac{s_V}{\sqrt{6}}\frac{1}{2s_Wc_W}
   \Big(\gamma_V^2-2s_W^2\frac{2(1-\gamma_V^2)}{3}\Big)\bigg)\phi_\mu
\nonumber\\&&
 +\bigg(\frac{s_V}{\sqrt{3}}\frac{1}{2s_Wc_W}
   \Big(\gamma_V^2-2s_W^2\frac{1+\gamma_V^2}{3}\Big)
 -\frac{c_V}{\sqrt{6}}\frac{1}{2s_Wc_W}
   \Big(\gamma_V^2-2s_W^2\frac{2(1-\gamma_V^2)}{3}\Big)\bigg)\omega_\mu
\Bigg)Z^\mu\Bigg\}
\nonumber\\&&
\frac{-iga}{2}
\Bigg\{\rho^{0,\mu}\bigg[\pi^+\hspace{-5pt}
      \stackrel{\leftrightarrow}{\partial}_\mu\hspace{-2pt}\pi^-
+\frac{\gamma_A^{-1}}{2}(1+\frac{C_A}{a}-C_V)
(K^+\hspace{-4pt}\stackrel{\leftrightarrow}{\partial}_\mu
      \hspace{-2pt}K^-
-K^0\hspace{-4pt}
      \stackrel{\leftrightarrow}{\partial}_\mu\hspace{-2pt}\bar{K}^0)\bigg]
\nonumber\\&&
+\phi^\mu\bigg[\bigg(\frac{c_V}{2\sqrt{3}}\Big(
 (\gamma_A^{-1}+2\gamma_V^2\gamma_A^{-1})
 +\frac{C_A}{a}(\gamma_A^{-1}+2)
 -C_V(\gamma_A^{-1}+2\gamma_V\gamma_A^{-1})\Big)
\nonumber\\&&
            -\frac{s_V}{\sqrt{6}}\Big(
 (\gamma_A^{-1}\hspace{-1.5pt}-\gamma_V^2\gamma_A^{-1})
 +\frac{C_A}{a}(\gamma_A^{-1}-1)
 -C_V(\gamma_A^{-1}\hspace{-1.5pt}-\gamma_V\gamma_A^{-1})\Big)\bigg)
\nonumber\\&&\times
   (K^+\hspace{-4pt}\stackrel{\leftrightarrow}{\partial}_\mu\hspace{-2pt}K^-
   \hspace{-3pt}+\hspace{-1pt}
   K^0\hspace{-4pt}\stackrel{\leftrightarrow}{\partial}_\mu\hspace{-2pt}
       \bar{K}^0)\bigg]
\nonumber\\&&
+\omega^\mu\bigg[\bigg(\frac{s_V}{2\sqrt{3}}\Big(
 (\gamma_A^{-1}+2\gamma_V^2\gamma_A^{-1})
 +\frac{C_A}{a}(\gamma_A^{-1}+2)
 -C_V(\gamma_A^{-1}+2\gamma_V\gamma_A^{-1})\Big)
\nonumber\\&&
            +\frac{c_V}{\sqrt{6}}\Big(
 (\gamma_A^{-1}-\gamma_V^2\gamma_A^{-1})
 +\frac{C_A}{a}(\gamma_A^{-1}-1)
 -C_V(\gamma_A^{-1}-\gamma_V\gamma_A^{-1})\Big)\bigg)
\nonumber\\&&\times
   (K^+\hspace{-4pt}\stackrel{\leftrightarrow}{\partial}_\mu\hspace{-2pt}K^-
   \hspace{-3pt}+\hspace{-1pt}
   K^0\hspace{-4pt}\stackrel{\leftrightarrow}{\partial}_\mu\hspace{-2pt}
\bar{K}^0)\bigg]
\nonumber\\&&
+K^{0*,\mu}\bigg[\gamma_V\gamma_A^{-\frac{1}{2}}
\bigg(-\frac{1}{\sqrt{2}}
 \pi^+\hspace{-5pt}\stackrel{\leftrightarrow}{\partial}_\mu\hspace{-2pt}K^-
+\frac{1}{2\sqrt{3}}(1+2\gamma_A^{-1})
 \bar{K}^0\hspace{-4pt}\stackrel{\leftrightarrow}{\partial}_\mu\hspace{-2pt}
        (c_P\eta+s_P\eta')
\nonumber\\&&
+\frac{1}{2} \pi^0\hspace{-4pt}\stackrel{\leftrightarrow}{\partial}_\mu\hspace{-2pt}
     \bar{K}^0
+\frac{1}{\sqrt{6}}(1-\gamma_A^{-\frac{1}{2}})
 \bar{K}^0\hspace{-4pt}\stackrel{\leftrightarrow}{\partial}_\mu
(-s_P\eta+c_P\eta')\bigg)
\nonumber\\&&
+(\frac{C_A}{a}-C_V)
\gamma_A^{-\frac{1}{2}}\bigg(\frac{1}{\sqrt{2}}K^-\partial_\mu\pi^+
 -\frac{1}{2}\bar{K}^0\partial_\mu\pi^0
 -\frac{2\gamma_A^{-1}}{\sqrt{3}}(c_P\eta+s_P\eta')\partial_\mu\bar{K}^0
\nonumber\\&&
 +\frac{2\gamma_A^{-1}}{\sqrt{6}}(-s_P\eta+c_P\eta')\partial_\mu\bar{K}^0
\bigg)
+\frac{C_A}{a}\frac{\sqrt{3}\gamma_A^{-\frac{1}{2}}}{2}
       \bar{K}^0\partial_\mu(c_P\eta+s_P\eta')
\nonumber\\&&
+C_V\bigg((-\frac{\gamma_V\gamma_A^{-\frac{3}{2}}}{\sqrt{3}}
      -\frac{\gamma_A^{-\frac{1}{2}}}{2\sqrt{3}})
           \bar{K}^0\partial_\mu(c_P\eta+s_P\eta') 
    +(\frac{\gamma_V\gamma_A^{-\frac{3}{2}}}{\sqrt{6}}
      -\frac{\gamma_A^{-\frac{1}{2}}}{\sqrt{6}})
\nonumber\\&&     
           \times\bar{K}^0\partial_\mu(c_P\eta'-s_P\eta)\bigg)\bigg]
\nonumber\\&&
+\bar{K}^{0*,\mu}\bigg[\gamma_V\gamma_A^{\frac{1}{2}}
\bigg(\frac{1}{\sqrt{2}}
 \pi^-\hspace{-5pt}\stackrel{\leftrightarrow}{\partial}_\mu\hspace{-2pt}K^+
-\frac{1}{2\sqrt{3}}(1+2\gamma_A^{-1})
 K^0\hspace{-4pt}\stackrel{\leftrightarrow}{\partial}_\mu\hspace{-2pt}
        (c_P\eta+s_P\eta')
\nonumber\\&&
-\frac{1}{2}
\pi^0\hspace{-4pt}\stackrel{\leftrightarrow}{\partial}_\mu\hspace{-2pt}
     K^0
-\frac{1}{\sqrt{6}}(1-\gamma_A^{-\frac{1}{2}})
K^0\hspace{-4pt}\stackrel{\leftrightarrow}{\partial}_\mu(-s_P\eta+c_P\eta')
\bigg)
\nonumber\\&&
+(\frac{C_A}{a}-C_V)
\gamma_A^{-\frac{1}{2}}\bigg(-\frac{1}{\sqrt{2}}K^+\partial_\mu\pi^-
 +\frac{1}{2}K^0\partial_\mu\pi^0
 +\frac{2\gamma_A^{-1}}{\sqrt{3}}(c_P\eta+s_P\eta')\partial_\mu K^0
\nonumber\\&&
 -\frac{2\gamma_A^{-1}}{\sqrt{6}}(-s_P\eta+c_P\eta')\partial_\mu K^0
\bigg)
-\frac{C_A}{a}\frac{\sqrt{3}\gamma_A^{-\frac{3}{2}}}{2}
       K^0\partial_\mu(c_P\eta+s_P\eta')
\nonumber\\&&
-C_V\bigg((-\frac{\gamma_V\gamma_A^{-\frac{3}{2}}}{\sqrt{3}}
      -\frac{\gamma_A^{-\frac{1}{2}}}{2\sqrt{3}})
           K^0\partial_\mu(c_P\eta+s_P\eta')
    +(\frac{\gamma_V\gamma_A^{-\frac{3}{2}}}{\sqrt{6}}
      -\frac{\gamma_A^{-\frac{1}{2}}}{\sqrt{6}})
\nonumber\\&&\times
           K^0\partial_\mu(-s_P\eta+c_P\eta')\bigg)\bigg]\Bigg\}
+\cdots
\end{eqnarray}
where $\gamma_{A,V}=C_{A,V}+1$ and $\pi$, $\eta$,\dots are renormalized
pseudoscalar fields (superscript $r$ is omitted). In the above expression, 
only the gauge-boson--vector-meson (first curly bracket) 
and vector-meson--two-pseudoscalar-meson (second curly bracket)
interactions are kept.

\setcounter{equation}{0}
\section{Form factors and loop functions}
\indent

The composite form factors for $\tau\to l'V^0$ decays, $a_{M^0}$,
$b_{M^0}$, and, $c_{M^0}$, appearing in the
first three Eqs.~(\ref{ABCDEF}), may be decomposed into the composite
loop form factors $F_\gamma^{\tau l'}$, $G_\gamma^{\tau l'}$,
$F_Z^{\tau l'}$, $F_{Box}^{\tau l'd_ad_b}$ and $F_{Box}^{\tau l'uu}$
in the following way
\begin{eqnarray}
a_{V^0}^{\tau
l'}&=&\frac{i\alpha^2_W}{16M_W^2}\frac{m_{V^0}^2}{\gamma_{V^0}}
\Big[\alpha_{V^0}^Z F_Z^{\tau l'}+\alpha_{V^0}^{Box,uu} F_{Box}^{\tau
l'uu}
+\alpha_{V^0}^{Box,dd} F_{Box}^{\tau l'dd}
\nonumber\\&&
+\alpha_{V^0}^{Box,ss} F_{Box}^{\tau l'ss}
+\alpha_{V^0}^{Box,ds} F_{Box}^{\tau l'ds}
+\alpha_{V^0}^{Box,sd} F_{Box}^{\tau l'sd}\Big],
\nonumber\\
b_{V^0}^{\tau
l'}&=&\frac{i\alpha^2_W}{16M_W^2}\frac{m_{V^0}^2}{\gamma_{V^0}}
\beta_{V^0}^\gamma\;F_\gamma^{\tau l'},
\nonumber\\
c_{V^0}^{\tau
l'}&=&\frac{i\alpha^2_W}{16M_W^2}\frac{m_{V^0}^2}{\gamma_{V^0}}
\gamma_{V^0}^\gamma\;G_\gamma^{\tau l'}.
\end{eqnarray}
The factors $\alpha_{V^0}^{\tau l'}$, $\beta_{V^0}^{\tau l'}$ and
$\gamma_{V^0}^{\tau l'}$, containing
information on quark content of a vector meson $V^0$
(see Table II), and in part
information on quark-$\gamma$ and quark-$Z^0$ couplings, may be
found in Table~I.

The loop form factors $F_\gamma^{\tau l'}$, $G_\gamma^{\tau l'}$,
$F_Z^{\tau l'}$, $F_{Box}^{\tau l'd_ad_b}$ and $F_{Box}^{\tau l'uu}$, 
and $F_H^{\tau l'}$ and $G_H^{\tau l'}$ contain the leptonic part of
$T_\gamma$, $T_Z$, $T_{Box}$ and $T_H$ amplitudes, and may be further
decomposed into elementary loop functions $F_\gamma$, $G_\gamma$,
$F_Z$, $G_Z$, $H_Z$, $F_{Box}$, $H_{Box}$, $F_H$, $G_H$, and $H_H$.
The loop form factors $F_\gamma^{\tau l'}$, $G_\gamma^{\tau l'}$,
$F_Z^{\tau l'}$, $F_{Box}^{\tau l'd_ad_b}$ and $F_{Box}^{\tau l'uu}$
together with the elementary loop functions $F_\gamma$, $G_\gamma$,
$F_Z$, $G_Z$, $H_Z$, $F_{Box}$, $H_{Box}$ may be found in Refs.~
\cite{NPB,PRD}. The composite loop form factor $G_H^{\tau l'}$ 
and the loop functions $F_H$ and $G_H$ were calculated for case of 
degenerate heavy neutrino masses in Ref.~\cite{AP_H}. Here the 
expressions for the composite loop formfactors 
$F_H^{\tau l'}$ and $G_H^{\tau l'}$ are listed
\begin{eqnarray}
F_H^{\tau l'}&=&\sum_{ij} B^\ast_{\tau i}B_{l'j}
              \bigg[\delta_{ij}F_H(\lambda_i)
              +C^\ast_{ij}G_H(\lambda_i,\lambda_j)
              +C_{ij}H_H(\lambda_i,\lambda_j)\bigg]
\nonumber\\
&=&\sum_{N_iN_j}B^\ast_{\tau N_i}B_{l'N_j} 
   \bigg[\delta_{N_iN_j}\Big(F_H(\lambda_{N_i})-F_H(0)
                        +G_H(\lambda_{N_i},0)+G_H(0,\lambda_{N_i})\Big)
\nonumber\\&&
        +C^\ast_{N_iN_j}\Big(G_H(\lambda_{N_i},\lambda_{N_j})
                        -G_H(\lambda_{N_i},0)-G_H(0,\lambda_{N_j})\Big)
        +C_{N_iN_j}H_H(\lambda_{N_i},\lambda_{N_j})\bigg],
\nonumber\\
G_H^{\tau l'}&=&\sum_{ij} B^\ast_{\tau i}B_{l'j}
              \bigg[\delta_{ij}F_H(\lambda_i)
              +C^\ast_{ij}G_H(\lambda_j,\lambda_i)
              +C_{ij}H_H(\lambda_j,\lambda_i)\bigg]
\nonumber\\
&=&\sum_{N_iN_j}B^\ast_{\tau N_i}B_{l'N_j} 
   \bigg[\delta_{N_iN_j}\Big(F_H(\lambda_{N_i})-F_H(0)
                        +G_H(\lambda_{N_i},0)+G_H(0,\lambda_{N_i})\Big)
\nonumber\\&&
        +C^\ast_{N_iN_j}\Big(G_H(\lambda_{N_j},\lambda_{N_i})
                        -G_H(\lambda_{N_j},0)-G_H(0,\lambda_{N_i})\Big)
        +C_{N_iN_j}H_H(\lambda_{N_j},\lambda_{N_i})\bigg],
\nonumber\\&&
\end{eqnarray}
together with the loop form factors $F_H$, $G_H$, and $H_H$ 
contained in them
\begin{eqnarray}
F_H(x)&=&
\frac{1-x+x\ln x}{(1-x)^2}
\bigg(\frac{x}{2}+\frac{x\lambda_H}{2}\bigg) 
+\bigg(\frac{3}{2}+\frac{x\ln x}{1-x}\bigg)
\frac{x}{2}
\nonumber\\&&
+\frac{1-4x+3x^2-2x^2\ln x}
{2(1-x)^3}\bigg(-\frac{3}{2}-\frac{x\lambda_H}{4}\bigg),
\nonumber\\
G_H(x,y)&=&
\frac{x(x-y)(1-x)(1-y)
     +x(1-y)(x+xy-2y)
      \ln x+y^2(1-x)^2\ln y}
     {-2(1-x)^2(1-y)(x-y)^2}
\nonumber\\&&
\times(x+y+xy)
+\frac{x\ln x-y\ln y
     -xy(\ln x-\ln y)}
     {(1-x)(1-y)(x-y)}y
\nonumber\\&&
+\bigg(-\frac{3}{4}
     +\frac{(1+x)\ln x-(1+y)\ln y}
           {2(x-y)}
     +\frac{1}{2(x-y)}
           \Big(\frac{\ln x}{-1+x}
              -\frac{\ln y}{-1+y}\Big)\bigg)y,
\nonumber\\
H_H(x,y)&=&
\sqrt{xy}\Bigg(
\frac{x(x-y)(1-x)(1-y)
     +x(1-y)(x+xy-2y)
      \ln x+y^2(1-x)^2\ln y}
     {-2(1-x)^2(1-y)(x-y)^2}
\nonumber\\&&
\times(2+\frac{1}{2}(x+y))
+\frac{x\ln x-y\ln y
     -xy(\ln x-\ln y)}
     {(1-x)(1-y)(x-y)}
\nonumber\\&&
+\bigg(-\frac{3}{4}
     +\frac{(1+x)\ln x-(1+y)\ln y}
           {2(x-y)}
     +\frac{1}{2(x-y)}
           \Big(\frac{\ln x}{-1+x}
              -\frac{\ln y}{-1+y}\Big)\bigg)\Bigg).
\nonumber\\&&
\end{eqnarray}
For reader's convenience, $F_H$, $G_H$, and $H_H$ are evaluated for some
special values of arguments,
\begin{eqnarray}
F_H(0)&=&-\frac{3}{4},\qquad F_H(1)=\frac{\alpha_H}{6},
\nonumber\\
G_H(x,x)&=&
\frac{-5x+4x^2+x^3-
      (10x^2-6x^3+2x^4)\ln x}
{4(1-x)^3},
\nonumber\\
G_H(x,1)&=&
\frac{-3+17x-13x^2-x^3+
      (14x^2-2x^3)\ln x}
{4(1-x)^3},
\nonumber\\
G_H(1,x)&=&
\frac{1-7x+8x^2-5x^3+3x^4
      -(6x^2-2x^3+2x^4)\ln x}
{4(1-x)^3},
\nonumber\\
G_H(x,0)&=&\frac{-x+x^2-x\ln x}
{2(1-x)^2},\qquad
G_H(0,x)=\frac{-3x+2x\ln x}{4},
\nonumber\\
G_H(1,1)&=&G_H(0,0)=0,\qquad G_H(0,1)=-\frac{3}{4},\qquad
G_H(1,0)=\frac{1}{4},
\nonumber\\
H_H(x,x)&=&
\frac{-5x+4x^2+x^3-
      (10x^2-6x^3+2x^4)\ln x}
{4(1-x)^3},
\nonumber\\
H_H(x,1)&=&
\frac{x^\frac{3}{2}(7-8x+x^2
      +(3+4x-x^2)\ln x)}
{2(1-x)^3},
\nonumber\\
H_H(1,x)&=&
\frac{x^\frac{1}{2}(-5+7x-11x^2+9x^3
      -(8x-2x^2+6x^3)\ln x)}
{8(1-x)^3},
\nonumber\\
H_H(0,x)&=&H_H(x,0)=H_H(1,1)=0.
\end{eqnarray}
If $s_L^{\nu_i}$ are kept constant, all composite 
loop form factors are increasing functions of the 
heavy neutrino masses. The asymptotic behaviour of the form factors
$F_\gamma^{\tau l'}$, $G_\gamma^{\tau l'}$ and $F_Z^{\tau l'}$, 
in the limit $\lambda_1\gg 1$ and
$\rho=\lambda_2/\lambda_1\geq 1$, are listed in Ref.~\cite{NPB}.
Here we list the form factors $F_H^{\tau l'}$ and $G_H^{\tau l'}$ 
in the same limit,
\begin{eqnarray}
F_H^{\tau l'},\; G_H^{\tau l'}&\to&s_L^{\nu_\tau}s_L^{\nu_{l'}}\:
 \Big(\frac{5}{8}+\frac{\lambda_H}{4}\ln \lambda_1
     +\frac{\lambda_H}{4}\frac{\ln\rho}{1+\rho^{1/2}}\Big)
\nonumber\\&&
 +s_L^{\nu_\tau}s_L^{\nu_{l'}}\sum_{l=1}^{n_G}(s_L^{\nu_l})^2\;
 \frac{3\rho \lambda_1(4+4\rho^{1/2}+(1-\rho^{1/2})\ln\rho)}
       {4(1+\rho^{1/2})^3}.
\end{eqnarray} 

\setcounter{equation}{0}
\section{Phase space functions}
\indent

The momentum dependent part of the absolute squares of the $\tau^-\to
l'^\mp P_1P_2$ amplitudes may be expressed in terms of the Mandelstam
variables $t=(p-p')^2$ and $s_1=(p'+p_1)^2=(p-p_2)^2$. The 
$\tau^-\to l'^\mp P_1P_2$ decay rates contain the integrals of the 
corresponding absolute squares of the amplitudes over $s_1$ and $t$ 
variables:
\begin{equation}
\label{decrat}
\Gamma(\tau^-\to l'^\mp P_1P_2)=
\frac{1}{256\pi^3m^3}\int_{(m_1+m_2)^2}^{(m-m')^2} dt\:
\int_{s_1^-}^{s_1^+}ds_1
\langle |T(\tau^-\to l'^\mp P_1P_2)|^2\rangle
\end{equation}
where $\langle {|T(\tau^-\to l'^\mp P_1P_2)|^2}\rangle$ is the 
square of the amplitude averaged over initial and summed over final
lepton spins. The boundary $s_1$-values, $s_1^\pm(t)$, read
\begin{equation}
\label{s1}
s_1^\pm(t)=m^2+m_2^2+\frac{B(t)}{A(t)}\pm\frac{\sqrt{B(t)^2-4A(t)C(t)}}
{A(t)},
\end{equation}
where
\begin{eqnarray}
\label{ABC}
A(t)&=&4t,\qquad B(t)\ =\ -2(m^2-m'^2+t)(t+m_2^2-m_1^2)
\nonumber\\
C(t)&=&m^2(t+m_2^2-m_1^2)^2+m_2^2\lambda(m^2,m'^2,t),
\end{eqnarray}
and $\lambda(x,y,z)=x^2+y^2+z^2-2xy-2xz-2yz$.
Since the momentum dependent parts of the squared amplitude
in Eq.~(\ref{decrat}) contain only powers of 
the $s_1$ variable, $s_1$ integration is
easily performed  resulting with expressions which are denoted by
$\alpha$, $\beta$, $\gamma$, $\delta$, $\varepsilon$, $\zeta$, $\eta$,
$\vartheta$, $\iota$, $\kappa$ and $\omega$:
\begin{eqnarray}
\label{alom}
\alpha &=& 2S_1^2+S_1^1\bigg[2t-2(m^2+m'^2+m_1^2+m_2^2)\bigg]
\nonumber\\&&
+S_1^0\bigg[-\frac{t}{2}(m^2+m'^2)+\frac{1}{2}(m^2+m'^2)^2+2m_1^2m_2^2\bigg],
\nonumber\\
\beta &=&S_1^1\bigg[m^2-m'^2\bigg]+S_1^0\bigg[\frac{t}{2}(m^2-m'^2)
-\frac{1}{2}(m^4-m'^4)-(m^2m_1^2-m'^2m_2^2)\bigg],
\nonumber\\
\gamma &=&S_1^0\bigg[-\frac{t}{2}(m^2+m'^2)+\frac{1}{2}(m^2-m'^2)^2\bigg],
\nonumber\\
\delta &=&S_1^1\bigg[\frac{1}{t}(m^2-m'^2)(m_1^2-m_2^2)\bigg]
+S_1^0\bigg[-\frac{t}{2}(m^2+m'^2)+\frac{1}{2}(m^2-m'^2)(m_1^2-m_2^2)
\nonumber\\&&
+\frac{1}{2}(m^2-m'^2)^2+(m_1^2+m_2^2)(m^2+m'^2)+\frac{1}{t}\bigg(
-\frac{1}{2}(m^4-m'^4)(m_1^2-m_2^2)
\nonumber\\&&
-(m^2m_1^2-m'^2m_2^2)(m_1^2-m_2^2)-(m_1^2+m_2^2)(m^2-m'^2)^2\bigg)\bigg],
\nonumber\\
\varepsilon &=&S_1^2\bigg[\frac{2}{t}(m^2+m'^2)\bigg]+S_1^1\bigg[2(m^2+m'^2)
-\frac{2}{t}\bigg((m^2+m'^2)^2+(m^2+m'^2)(m_1^2+m_2^2)\bigg)
\nonumber\\&&
-\frac{2}{t^2}(m^4-m'^4)(m_1^2-m_2^2)\bigg]+S_1^0\bigg[\frac{t}{2}
(m^2+m'^2)-\frac{1}{2}(m^2-m'^2)^2
\nonumber\\&&
-(m^2+m'^2)(m_1^2+m_2^2)
+\frac{1}{t}\bigg(2m^2m'^2(m^2+m'^2)-4m^2m'^2(m_1^2+m_2^2)
\nonumber\\&&
-(m^4-m'^4)(m_1^2-m_2^2)
+\frac{1}{2}(m^2+m'^2)(m_1^2+m_2^2)^2\bigg)
\nonumber\\&&
+\frac{1}{t^2}\bigg(2(m^4-m'^4)(m^2m_1^2-m'^2m_2^2)+(m^4-m'^4)(m_1^4-m_2^4)
\nonumber\\&&
+(\frac{1}{2}m^4+3m^2m'^2+\frac{1}{2}m'^4)(m_1^2-m_2^2)^2\bigg)\bigg]
\nonumber\\
\zeta &=&m\: 
\bigg(S_1^1+S_1^0\bigg[\frac{t}{2}-\frac{1}{2}(m^2+m'^2)
-m_1^2\bigg]\bigg),
\nonumber\\
\eta &=&m\:
S_1^0\:\bigg[-\frac{t}{2}+\frac{m^2-m'^2}{2}\bigg],
\nonumber\\
\vartheta &=&m\: \bigg(
S_1^1\bigg[\frac{1}{t}(m^2-m'^2)\bigg]+
S_1^0\bigg[-\frac{1}{2t}(m^2-m'^2)(m^2+m'^2+m_1^2+m_2^2)
\nonumber\\&&
+\frac{1}{2}(m^2-m'^2+m_2^2-m_1^2)\bigg]\bigg),
\nonumber\\
\iota &=&S_1^0\: \bigg[\frac{1}{2}(m^2+m'^2-t)\bigg],
\nonumber\\
\kappa &=&mm'\: S_1^0,
\nonumber\\
\omega &=&S_1^0\: \bigg[\frac{1}{2}(t-m_1^2-m_2^2)^2(m^2+m'^2-t)\bigg],
\end{eqnarray}
where
\begin{equation}
\label{S1n}
S_1^n=\int_{s_1^-(t)}^{s_1^+(t)}ds_1\:s_1^n.
\end{equation}
The definitions of other quantities in Eqs.~
(\ref{decrat}--\ref{ABC}) may be found in the previous
text. The t-integration of expressions (\ref{decrat}) has been performed
numerically.
\end{appendix}

\newpage

\centerline{\Large\bf Figure Captions}

\newcounter{fig}
\begin{list}{\bf\rm Fig.~\arabic{fig}: }{\usecounter{fig}
\labelwidth1.6cm \leftmargin2.5cm \labelsep0.4cm \itemsep0ex plus0.2ex }

\item Feynman graphs pertinent to the semileptonic
lepton-number-violating decays $\tau^-\to l'^+P_1^-P_2^-$ [Fig.1(a)]
and to the semileptonic lepton-flavour violating decays 
$\tau^-\to l'^-P_1P_2$ [Fig.1(b)]. The hatched blobs represent sets of
lowest order diagrams contributing to three-point and four-point
functions violating lepton flavour. These sets of diagrams may be found
in Refs. \cite{NPB,PRD,AP_H,AP_pertu,AP}. The double hatched blobs
represent interactions through which the final state pseudoscalar mesons
are formed.

\item Branching ratios (BR-s) versus heavy-neutrino mass
$m_N=m_{N_1}=\frac{1}{3}m_{N_2}$ for the decays 
$\tau^-\to e^-\pi^-\pi^+$ (thick solid line),
$\tau^-\to e^-K^-K^+$ (thick dashed line),
$\tau^-\to e^-K^0\bar{K}^0$ (thick dot-dashed line),
$\tau^-\to e^-\pi^-K^+/e^-\pi^+K^-$ (1),
$\tau^-\to e^-\pi^0K^0/e^-\pi^0\bar{K}^0$ (2),
$\tau^-\to e^-\eta K^0/e^-\eta \bar{K}^0$ (3),
$\tau^-\to e^-\eta' K^0/e^-\eta' \bar{K}^0$ (4),
$\tau^-\to e^-\pi^0\pi^0$ (5),
$\tau^-\to e^-\eta\eta$ (6),
$\tau^-\to e^-\eta\eta'$ (7),
$\tau^-\to e^+\pi^-\pi^-$ (8),
$\tau^-\to e^+\pi^-K^-$ (9),
$\tau^-\to e^+K^-K^-$ (10),
assuming $(s^{\nu_e}_L)^2=0.01$ and $(s^{\nu_\tau}_L)^2=0.05$.

\item Branching ratios versus new electroweak parameters of the 
model.
Fig.~3(a): BR-s versus $m_N=m_{N_1}=m_{N_2}$, assuming
$(s^{\nu_e}_L)^2=0.01$ and $(s^{\nu_\tau}_L)^2=0.05$.
Fig.~3(b): BR-s versus $m_N=m_{N_1}=m_{N_2}$, assuming
$(s^{\nu_e}_L)^2=0.01$ and $(s^{\nu_\tau}_L)^2=0.02$.
Fig.~3(c): BR-s versus $(s^{\nu_\tau}_L)^2$, assuming
$m_N\ =\ 4000\ GeV$ and $(s^{\nu_e}_L)^2=0.01$.
Fig.~3(d): BR-s versus $(s^{\nu_e}_L)^2$, assuming
$m_N\ =\ 4000\ GeV$ and $(s^{\nu_\tau}_L)^2=0.05$.

\item Branching ratios versus ratio $m_{N_2}/m_{N_1}$ 
for the decays of Fig. 3,
assuming $m_{N_1}=m_{N_2}=4$~TeV, $(s^{\nu_e}_L)^2=0.01$ and
$(s^{\nu_\tau}_L)^2=0.05$.

\item Branching ratios versus $m_N=m_{N_1}=m_{N_2}$ for 
the decays of Fig.~3, assuming $(s^{\nu_e}_L)^2\ =\ 0.01$ and
$(s^{\nu_\tau}_L)^2=0.05$. The figure illustrates the dependence of 
BR-s on few ingredients of hadronic part of the amplitudes.
 Fig.~5(a): The influence
of the vector meson propagators on BR-s. Fig.~5(b): The
influence of the $U(3)_L\times U(3)_R/U(3)_V$ breaking 
on BR-s. Fig.~5(c): BR-s for $\theta_V\ =\ 30^\circ$.
Thin lines represent the reference graphs and coincide with
thick lines in the Fig.~3(a). Thick lines show BR-s in a situation
when one of the ingredients of the hadronic part of the amplitudes is
changed.

\item Partial decay rates devided by the $\tau$ decay width as 
functions of $t=(p-p')^2$ assuming $m_{N_1}\ =\ m_{N_2}\ =\ 3700\ GeV$,
$(s^{\nu_e}_L)^2=0.01$ and $(s^{\nu_\tau}_L)^2=0.05$.

\end{list}

\newpage

\noindent
{\bf Table~I:} Coefficients defining composite form factors for $\tau\to
l'\bar{V}^0$ decays:\\
Besides the constants listed in the Table~I,
there are two more constants different from zero:
$\alpha_{K^{0*}}^{Box,ds}=\frac{1}{\sqrt{2}}$ and
$\alpha_{\bar{K}^{0*}}^{Box,ds}=\frac{1}{\sqrt{2}}$.
\\ \\
\begin{tabular}{|l|llllll|}\hline
$V^0$ & $\alpha_{V^0}^Z$ & $\alpha_{V^0}^{Box,uu}$ 
      & $\alpha_{V^0}^{Box,dd}$ & $\alpha_{V^0}^{Box,ss}$  
      & $\beta_{V^0}^\gamma$ & $\gamma_{V^0}^\gamma$ \\ \hline
$\rho^0$& $c_{2W}$ & $\frac{1}{2}$ & $\frac{1}{2}$ & $0$ & 
        $2s_W^2$ & $-2s_W^2$ \\
$\omega$& $\frac{s_Vc_{2W}}{\sqrt{3}}-\frac{c_V}{\sqrt{6}}$ 
      & $\frac{s_V}{2\sqrt{3}}+\frac{c_V}{\sqrt{6}}$ 
      & $-\frac{s_V}{2\sqrt{3}}-\frac{c_V}{\sqrt{6}}$ 
      & $\frac{s_V}{\sqrt{3}}-\frac{c_V}{\sqrt{6}}$
      & $\frac{2}{\sqrt{3}}s_W^2s_V$ 
      & $-\frac{2}{\sqrt{3}}s_W^2s_V$ \\
$\phi$& $\frac{c_Vc_{2W}}{\sqrt{3}}+\frac{s_V}{\sqrt{6}}$ 
      & $\frac{c_V}{2\sqrt{3}}-\frac{s_V}{\sqrt{6}}$
      & $-\frac{c_V}{2\sqrt{3}}+\frac{s_V}{\sqrt{6}}$
      & $\frac{c_V}{\sqrt{3}}+\frac{s_V}{\sqrt{6}}$
      & $\frac{2}{\sqrt{3}}s_W^2c_V$
      & $-\frac{2}{\sqrt{3}}s_W^2c_V$ \\ \hline
\end{tabular}
\\ \\ \\ \\
\noindent
{\bf Table~II:} Quark content of the pseudoscalar meson states and
fields:\\
The meson states listed in the Table~II correspond to the tensor
description of meson states, what is more appropriate for chiral model
calculations. The states $|\pi^+\rangle$ and $\bar{K}^0\rangle$ have
opposite signs from that refered in Ref. \cite{PRD}.
\\ \\
\hspace{1cm}
\begin{tabular}{|l|l|l|}\hline
$|M\rangle$ & quark content of $|M\rangle$ & quark content of $M(x)$\\
\hline
$|K^+\rangle$ & $us^c\sim b^\dagger_ud^\dagger_s$ & $su^c\sim d_sb_u$\\
$|K^0\rangle$ & $ds^c$ & $sd^c$\\
$|\pi^+\rangle$ & $ud^c$ & $du^c$\\
$|\pi^0\rangle$ & $\frac{1}{\sqrt{2}}(uu^c-dd^c)$ &
    $\frac{1}{\sqrt{2}}(uu^c-dd^c)$\\
$|\pi^-\rangle$ & $du^c$ & $ud^c$\\
$|K^-\rangle$ & $su^c$ & $us^c$\\
$|\bar{K}^0\rangle$ & $sd^c$ & $ds^c$\\
$|\eta_8\rangle$ & $\frac{1}{\sqrt{6}}(uu^c+dd^c-2ss^c)$ &
    $\frac{1}{\sqrt{6}}(uu^c+dd^c-2ss^c)$\\
$|\eta_1\rangle$ & $\frac{1}{\sqrt{6}}(uu^c+dd^c+ss^c)$ &
    $\frac{1}{\sqrt{6}}(uu^c+dd^c+ss^c)$\\ \hline
$|\eta\rangle$ & $\cos\theta_P|\eta_8\rangle
-\sin\theta_P|\eta_1\rangle$ &
    $\cos\theta_P\eta_8(x)-\sin\theta_P\eta_1(x)$\\
$|\eta'\rangle$ & $\sin\theta_P|\eta_8\rangle
+\cos\theta_P|\eta_1\rangle$ &
    $\sin\theta_P\eta_8(x)+\cos\theta_P\eta_1(x)$\\ \hline
\end{tabular}
\\
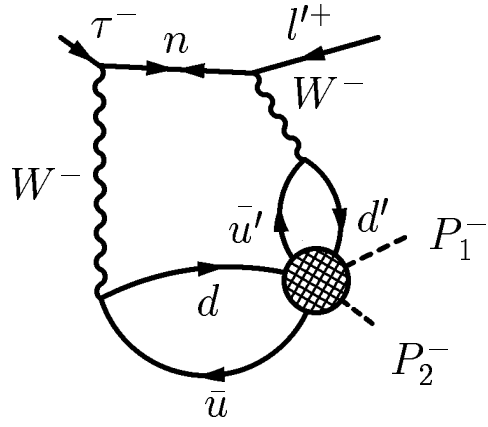
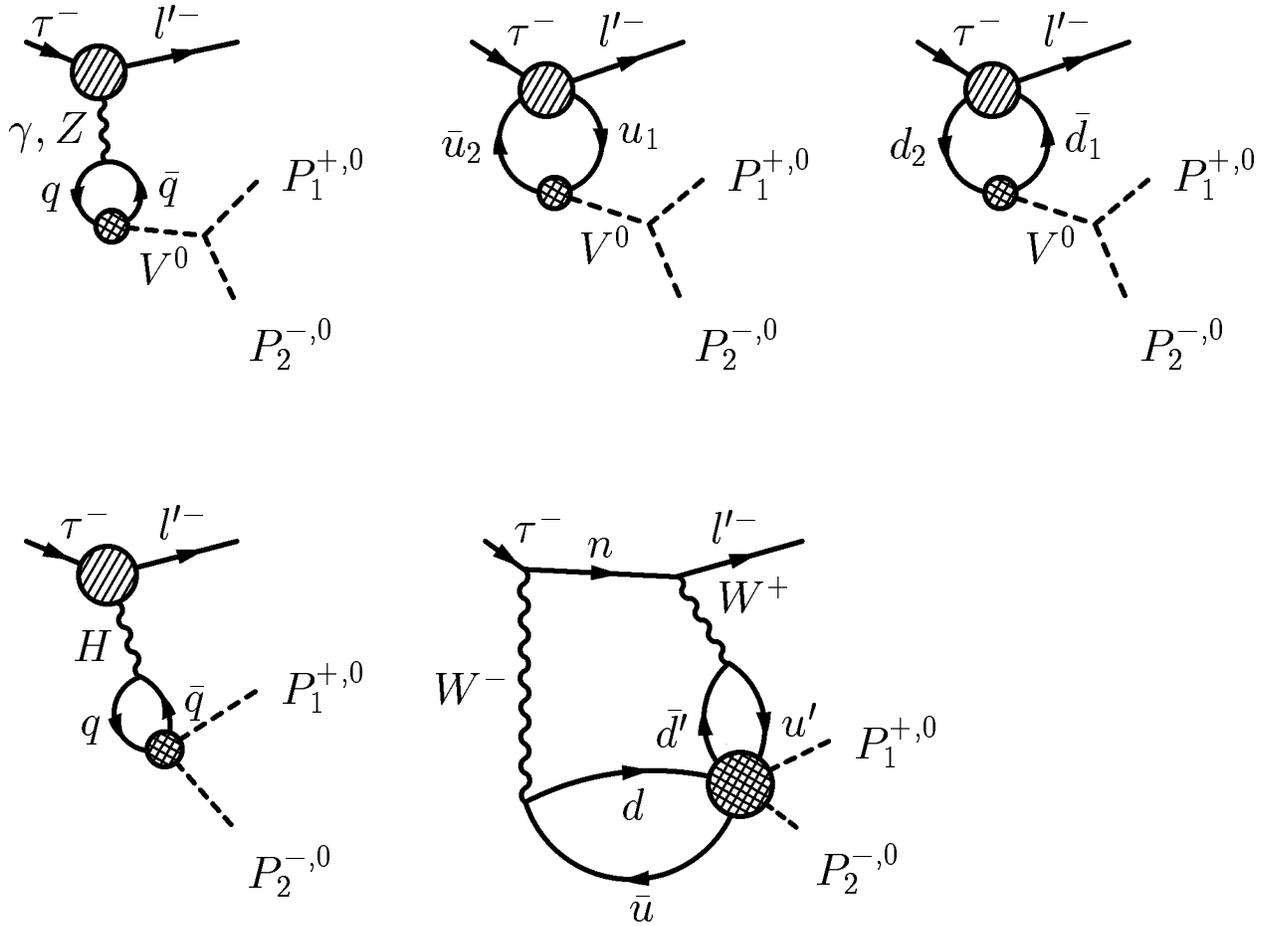
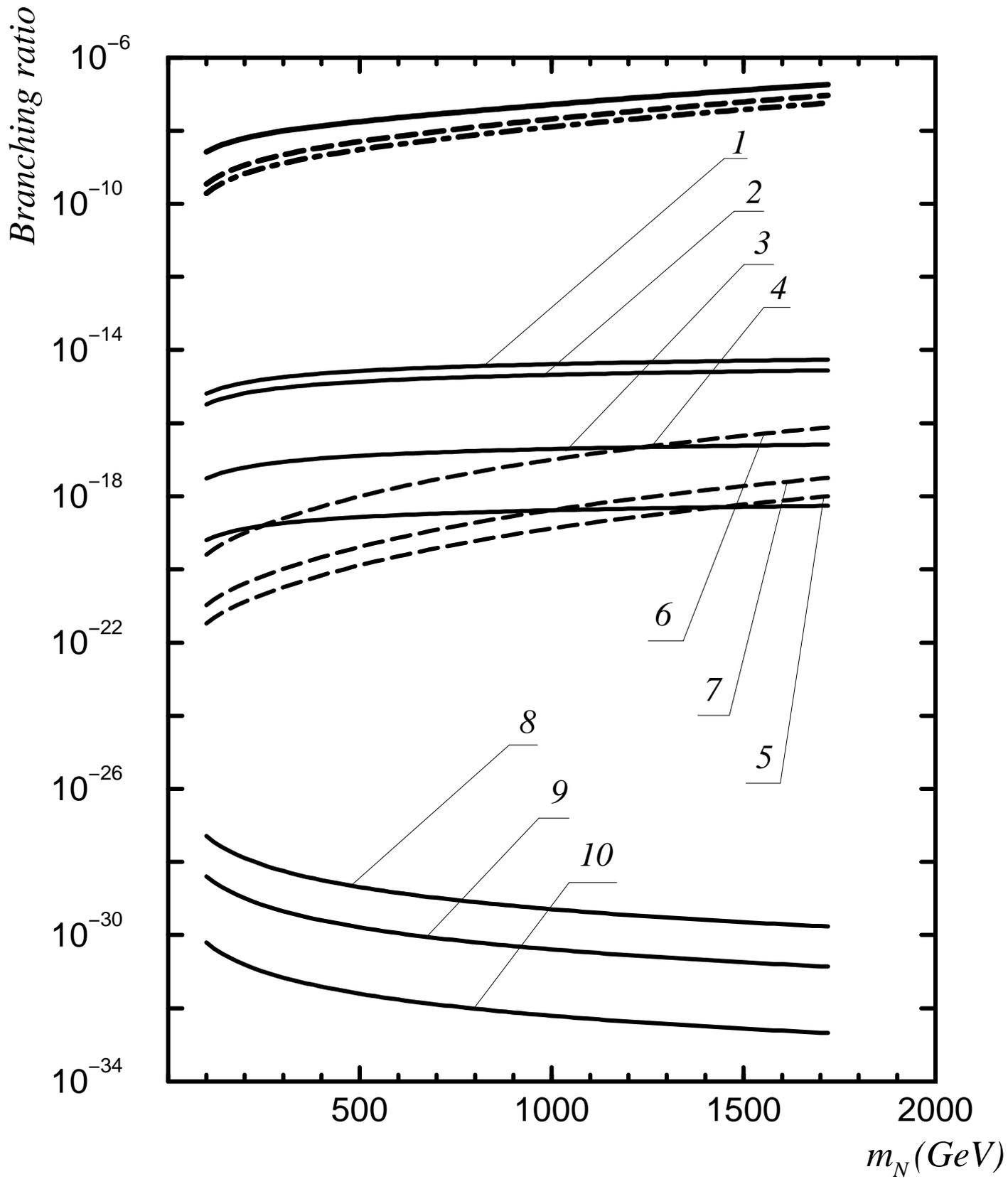
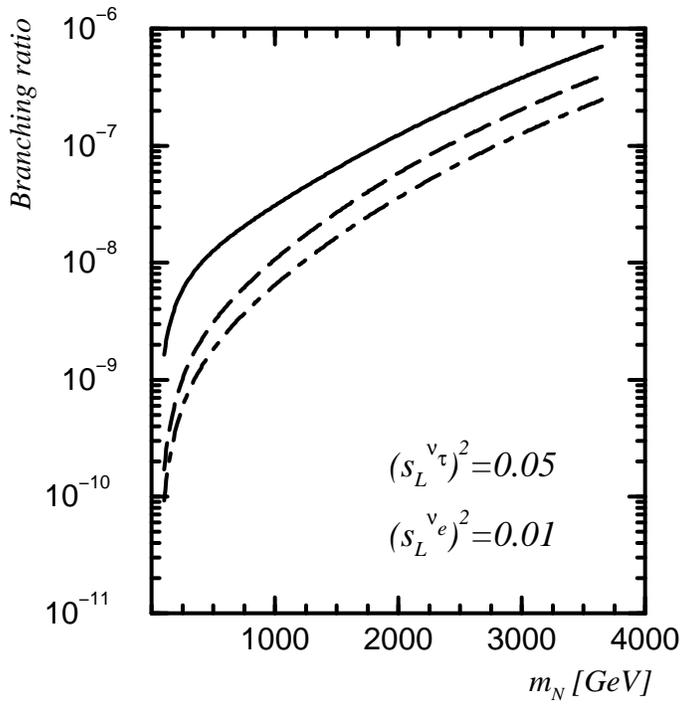
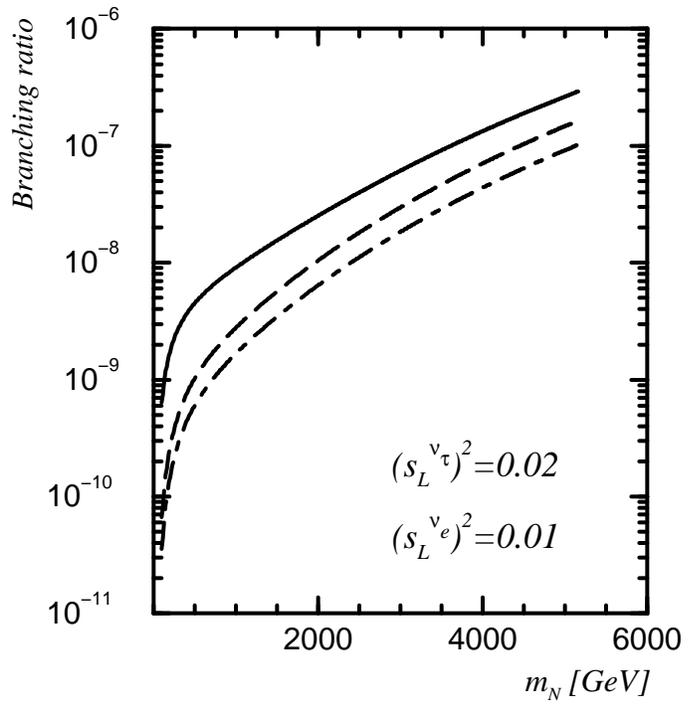
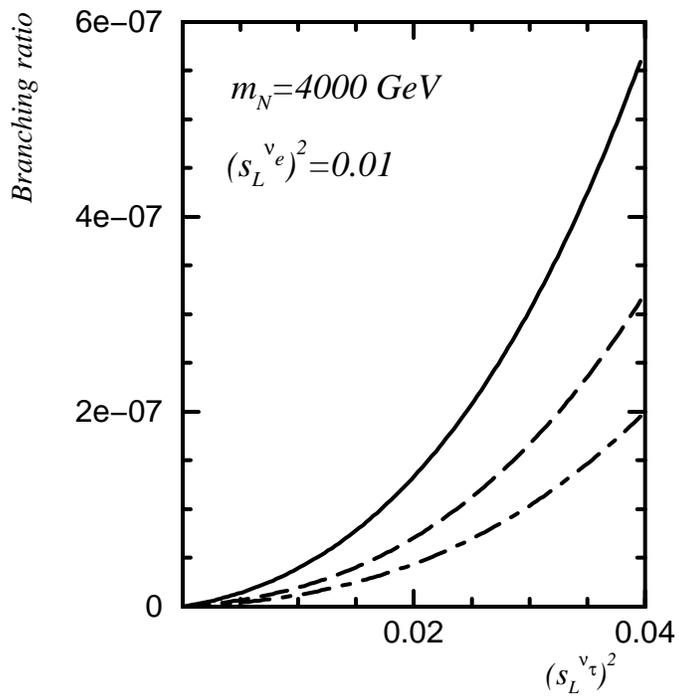
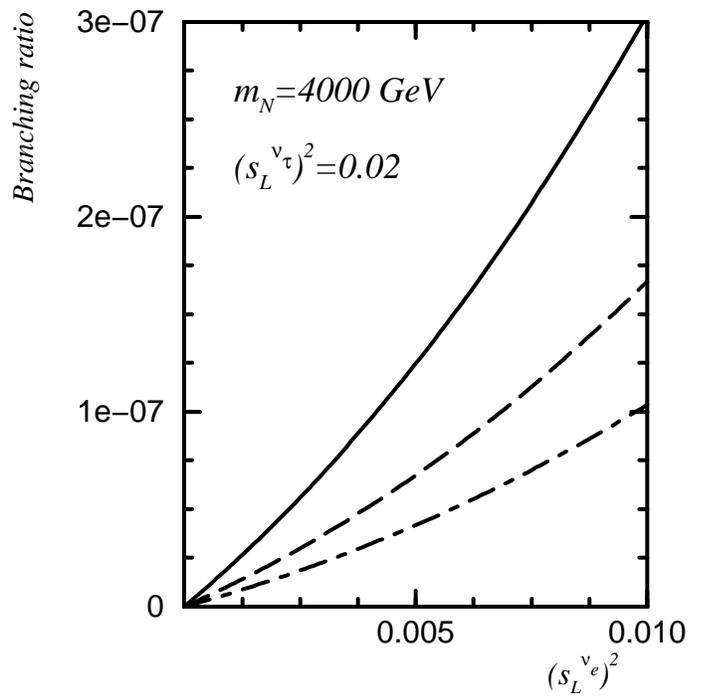
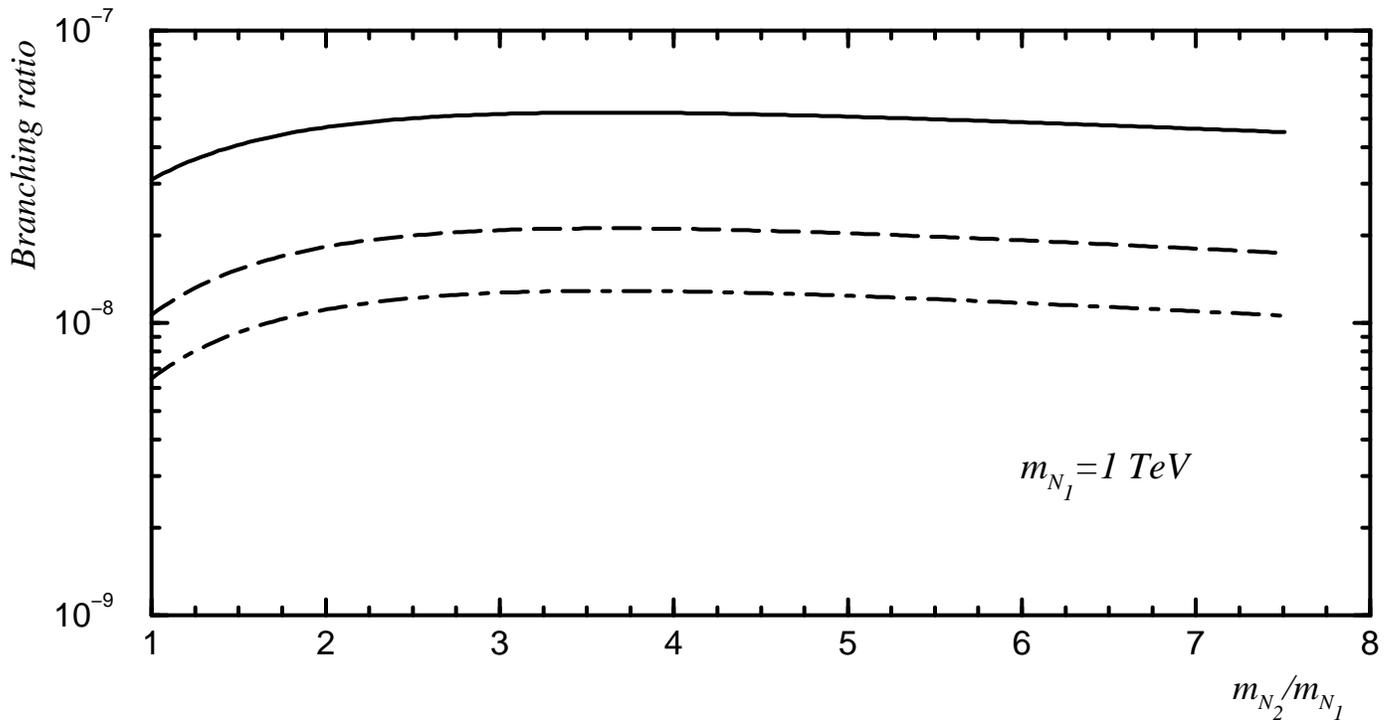
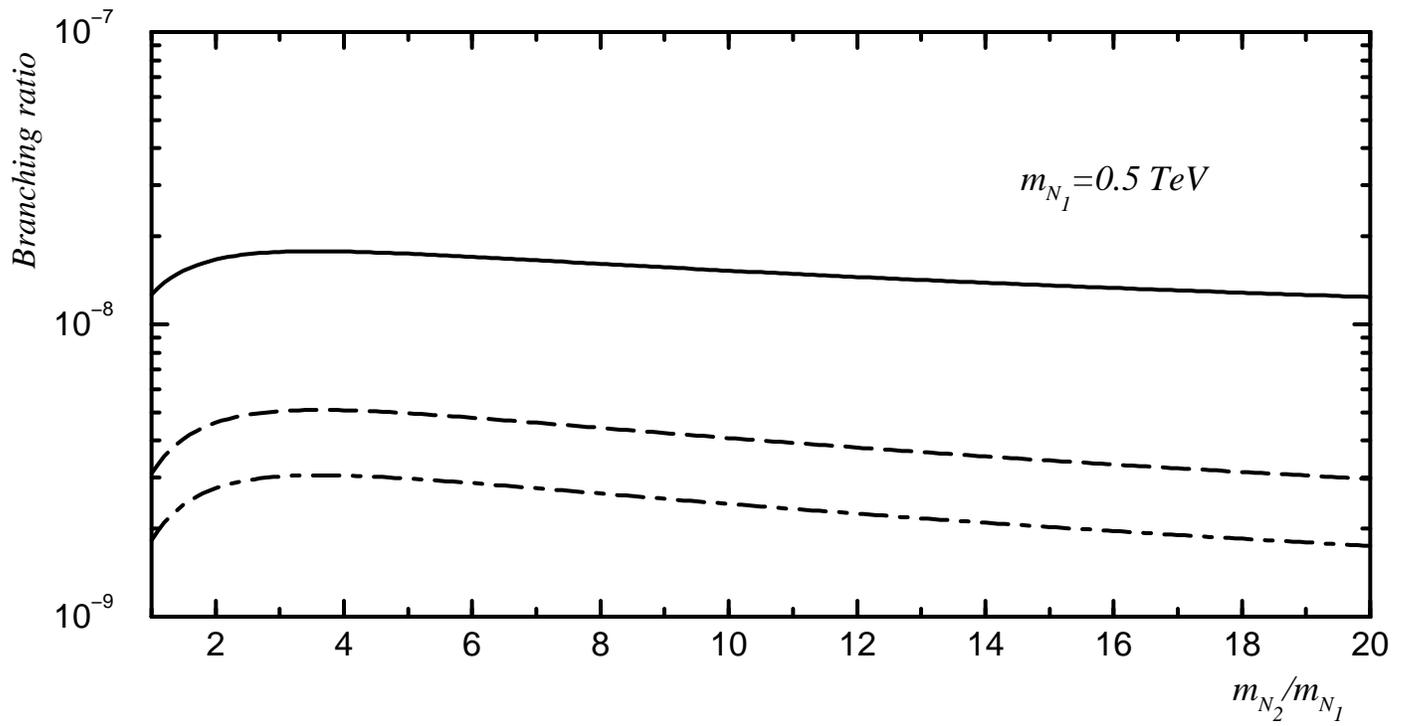
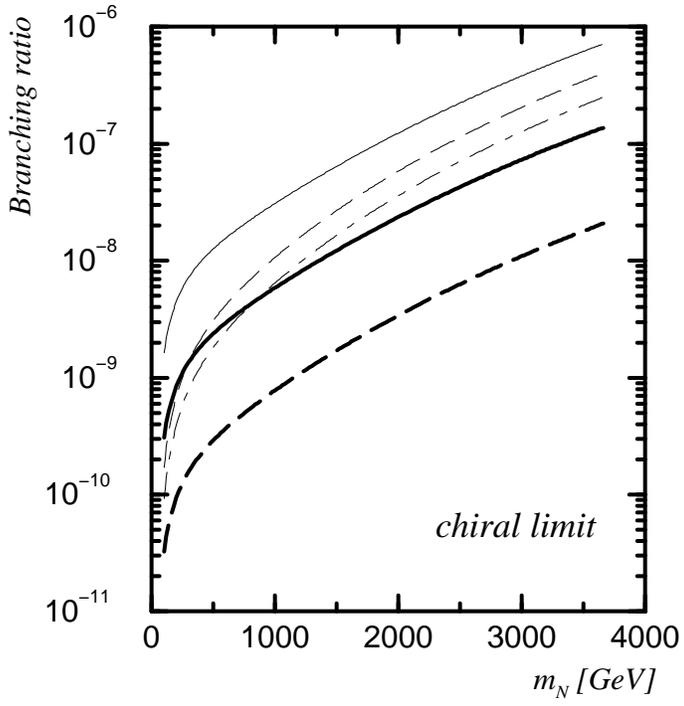
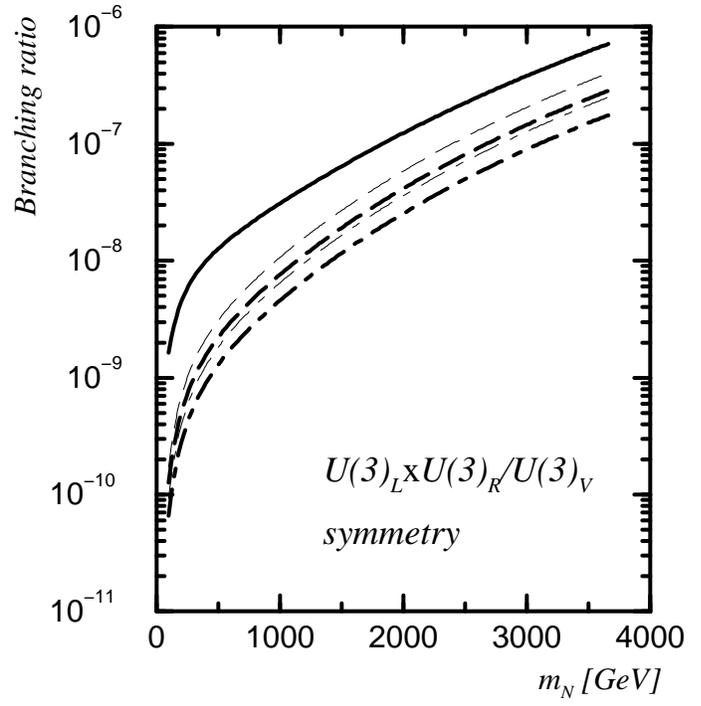
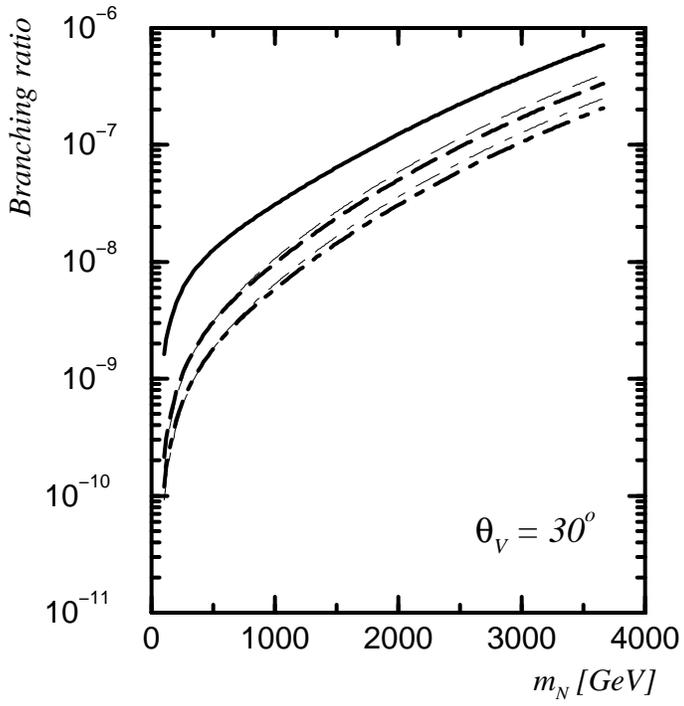
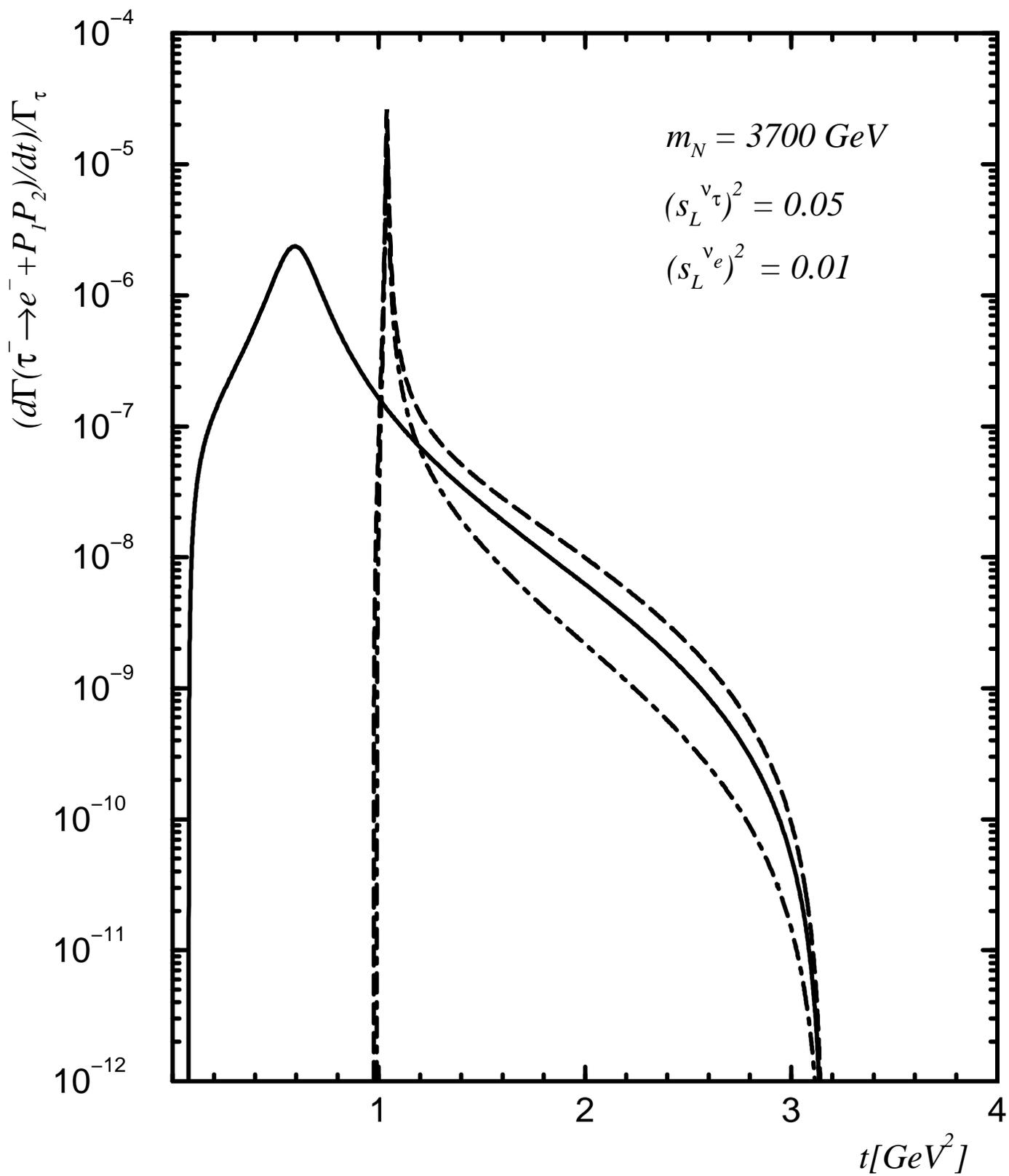
\end{document}